\documentclass[twocolumn]{aastex62}
\usepackage[utf8]{inputenc}

\usepackage{subfiles}
\usepackage{color}
\usepackage{longtable}
\usepackage{multirow}
\usepackage{lineno}
\graphicspath{{./}{Figures/}}

\defcitealias{edwards_ariel_tl}{E19}
\defcitealias{barclay_2018}{B18}
\defcitealias{barclay_2020}{B20}

\received{December 28, 2021}
\revised{March 25, 2022}
\accepted{April 28, 2022}

\submitjournal{AJ}

\shorttitle{The Ariel Target List}
\shortauthors{Edwards \& Tinetti}

\begin{document}

\title{The Ariel Target List: The Impact of TESS and the Potential for Characterising Multiple Planets Within a System}

\correspondingauthor{Billy Edwards}
\email{billy.edwards.16@ucl.ac.uk}

\author[0000-0002-5494-3237]{Billy Edwards}\thanks{Paris Region Fellow}
\affil{AIM, CEA, CNRS, Universit\'e Paris-Saclay, Universit\'e de Paris, F-91191 Gif-sur-Yvette, France}
\affil{Department of Physics and Astronomy, University College London, Gower Street, London, WC1E 6BT, UK}

\author[0000-0001-6058-6654]{Giovanna Tinetti}
\affil{Department of Physics and Astronomy, University College London, Gower Street, London, WC1E 6BT, UK}

\begin{abstract}

The ESA Ariel mission has been adopted for launch in 2029 and will conduct a survey of around one thousand exoplanetary atmospheres during its primary mission life. By providing homogeneous datasets, with a high SNR and wide wavelength coverage, Ariel will unveil the atmospheric demographics of these far-away worlds, helping to constrain planet formation and evolution processes on a galactic scale. Ariel seeks to undertake a statistical survey of a diverse population of planets and, therefore, the sample of planets from which this selection can be made is of the utmost importance. While many suitable targets have already been found, hundreds more will be discovered before the mission is operational. Previous studies have used predictions of exoplanet detections to forecast the available planet population by the launch date of Ariel, with the most recent noting that the Transiting Exoplanet Survey Satellite (TESS) alone should provide over a thousand potential targets. In this work, we consider the planet candidates found to date by TESS to show that, with the addition of already confirmed planets, Ariel will already have a more than sufficient sample to choose its target list from once these candidates are validated. We showcase the breadth of this population as well as exploring, for the first time, the ability of Ariel to characterise multiple planets within a single system. Comparative planetology of worlds orbiting the same star, as well as across the wider population, will undoubtedly revolutionise our understanding of planet formation and evolution.\\

\end{abstract}

\section{Introduction}

Ariel has been selected as the next ESA medium-class science mission and is due for launch in 2029 \citep{tinetti_ariel2}. During its 4-year mission, Ariel will observe 1000 exoplanet atmospheres, aiming to provide a diverse catalogue of homogeneous datasets which allow for the large-scale demographics of exoplanet atmospheres to be uncovered for the first time. The planets studied will span a wide range of planetary and stellar parameters, allowing Ariel to probe all corners of the exoplanet population, from temperate terrestrials to ultra-hot Jupiters. Data from Ariel will reveal the chemical fingerprints of gases and condensates in the planets' atmospheres, including the elemental composition and thermal structure. 

Ariel will simultaneously provide spectral coverage from 0.5-7.8\,$\mu m$, with photometric bands covering the visible and spectrometers providing data at wavelengths longer than 1.1\,$\mu m$. The mission objective of Ariel is to uncover the chemical diversity of exoplanet atmospheres, with the bulk of the mission being dedicated to a survey constructed of three Tiers where the depth to which the planet is studied increases with each tier. 

The list of potential targets for Ariel has been rapidly evolving over recent years and will continue to do so over the period until its launch. The evolution is driven by the plethora of planet detection surveys, each bringing unique parameter spaces to the fore and thus providing a multifarious population from which to select atmospheric targets. Ground-based surveys \citep[e.g.][]{bakos,pollacco,wheatley_ngts} have been instrumental in proving the success of the transit detection method, providing a multitude of hot gaseous planets as well as cooler, rocky worlds, while the Convection, Rotation and planetary Transits \citep[CoRoT,][]{auvergne_corot} mission was the first to detect exoplanets from space. The Kepler mission \citep{borucki_kepler} has provided the majority of known transiting planets to date, with the extended mission bringing yet more with a focus on brighter stars along the ecliptic \citep{howell_k2}. Finally, the Transiting Exoplanet Survey Satellite \citep[TESS,][]{ricker} has been operational since July 2018 and has been predicted to find thousands of planets \citep{sullivan,barclay_2018}, many of which will be suitable for atmospheric characterisation with Ariel and other facilities.

Here, we build upon other works which have explored the potential target list for Ariel \citep{zingales,edwards_ariel_tl}. The most recent of these, \citet{edwards_ariel_tl}, hereafter \citetalias{edwards_ariel_tl}, was conducted as the TESS mission launched and utilised the predictions of \citet{barclay_2018}, hereafter \citetalias{barclay_2018}, in addition to the planets known at the time, to projected the expected numbers of planets that could be studied during the mission's primary life. In the summer of 2020, TESS completed its primary 2-year mission and moved into an extended operations, resurveying parts of the northern and southern hemisphere, as well as covering the ecliptic plane. An updated study of the TESS planet yield was published \citet{barclay_2020}, hereafter \citetalias{barclay_2020}, suggesting that the extended mission would discover hundreds of additional worlds. Over the last few years, thousands of potential planet signals have been found within TESS data \citep{guerrero_tois} with 205 planets having been subsequently confirmed to date\footnote{NASA Exoplanet Archive, accessed 26th April 2022}.

In this work, we compare the current TESS TOIs to the detections predicted by \citetalias{barclay_2018} and \citetalias{barclay_2020}, with a specific focus on those which are suitable for study with Ariel. We explore not only the number of TOIs Ariel could study, but the variety in their properties. In \citetalias{edwards_ariel_tl}, a focus was placed upon Ariel's capabilities to study smaller, potential rocky worlds which could host secondary atmospheres. Here we focus instead on the systems in which there are multiple planets which are suitable for study with Ariel, highlighting the mission's great potential for comparative planetology within a single planetary system as well as across the vast exoplanet population. Finally, we discuss the efforts that are required to ensure a robust list of potential target is chosen for observation with Ariel.


\section{Construction of Catalogues}

\subsection{Currently-Known Planets and Those Predicted to be Found By TESS}

The catalogue of currently-known targets was built in an identical fashion to that of \citetalias{edwards_ariel_tl} although it was updated to include the detections that have occurred in the intervening years. As described in \citetalias{edwards_ariel_tl}, the catalogue is built mainly from the NASA Exoplanet Archive\footnote{\url{https://exoplanetarchive.ipac.caltech.edu/}} \citep{akeson_archive}. Since \citetalias{edwards_ariel_tl}, the NASA Exoplanet Archive has been updated to include a ``Planetary Systems Composite Data" table. The table ensures that as many planetary and stellar properties as possible are reported for a system which can mean the data is gathered from multiple studies. Both tables were accessed on 26$^{\rm th}$ April 2022. Where certain parameters were not available, we made further attempts to infer them. These included inferring the stellar mass, radius or temperature from  \citet{pecaut_stars}\footnote{\url{http://www.pas.rochester.edu/~emamajek/EEM_dwarf_UBVIJHK_colors_Teff.txt}}. If the mass had been measured in multiple studies, we took the value from the latest work, assuming that to be the most accurate. If it had not been measured, we estimated it using the relation from \citet{chen}. We removed any planets which did not have reported uncertainties on the planet radius. Having inferred as many parameters as possible, we also removed any planets which did not have the information required by Ariel instrument simulator, ArielRad \citep{mugnai_ar}. For the host star, these are the star radius, temperature, log(g) and distance. For the planet, the radius, mass, temperature and transit duration are needed. Having removed those with insufficient information, the input population consisted of 3488 planets.

Similarly, the \citetalias{barclay_2018} sample was the same as that considered in \citetalias{edwards_ariel_tl}, although the phantom inflated planets that were identified within the sample by  \citet{mayorga_pip} were removed, reducing the number of predicted planets to 4231. The yield from \citetalias{barclay_2020} was not considered in \citetalias{edwards_ariel_tl} but we include those planets here as an additional means of bench-marking the performance of TESS. These planets were treated in an identical way to those from \citetalias{barclay_2018} and the performance of Ariel was modelled for 7603 of them in this study.

\subsection{TESS Objects of Interest}

We accessed the latest TOIs from both the NASA Exoplanet Archive and the NASA ExoFOP-TESS site\footnote{\url{https://tev.mit.edu/data/}}, splicing them together using the TOI ID. Both lists were accessed on 26$^{\rm th}$ April 2022. The TOI list contains objects which have been flagged as false positives, such as eclipsing binaries, or as previously known planets \citep{guerrero_tois}. We used the ``TOI Disposition" column to determine the current status of a target. We filtered out known planets which were denoted by ``KP''. However, we found that, in the ``Public Comments" column, references to know systems were still made. Therefore, having made all comments lower case, we searched this column for the following keywords to identify known planets: `corot', `hat', `hd', `hip', `gj', `kelt', `kepler', `k2', `known', `lhs', `mascara', `ogle', `qatar' `tres', `wasp' and `xo'. Furthermore, we searched this same column for `eb' to filter out those identified as eclipsing binaries as well as comments which contained the phrase `v-shaped'. To avoid double counting planets in our analysis, we removed those labelled ``CP" in the ``TOI Disposition" clumn as these are denoting TOIs which have since gone on to be confirmed planets from the TESS mission. Removing all those that were flagged by the above process, resulted in 4192 remaining TOIs from the initial sample of 5637. Our acceptance rate is consistent with the $\sim$25\% false positive rate found by \citet{guerrero_tois} for the prime mission.



As mentioned, ArielRad, the instrument simulator used to model the performance of Ariel's photometers and spectrometers \citep{mugnai_ar}, requires a number of input parameters. Some of these were not given in the TOI list and so these were inferred using the methods of \citetalias{edwards_ariel_tl}. For instance, the planet mass is obviously not given in the TOI list as it is only known after further follow-up has been conducted. Hence, we used Forecaster \citep{chen} to estimate the mass. Additionally, we note that an estimate of the planet's temperature is given within the TOI list but we recalculate it to ensure compatibility between these targets and the predicted targets from \citetalias{barclay_2018}. Having followed this procedure, 3697 TOIs had enough information to be fed into ArielRad. From this point forward, we refer to these as TESS Planet Candidates (TPCs).

\section{Potential Candidates for Atmospheric Study with Ariel}

Ariel aims to undertake a meticulous chemical survey, searching for trends in atmospheric composition and unveiling the demographics of exoplanet atmospheres. Planning of observations with Ariel is based around a tiered approach which we briefly summarise here. Initially $\sim$1000 planets will be studied with the resulting spectra providing a basic characterisation of the atmosphere (presence of clouds, colour-colour diagrams etc.). From this sample, around half will be studied in more depth, with additional time dedicated to them to build up the SNR of the spectra. These Tier 2 observations will provide a more detailed view of the atmosphere, constraining the trace gases, metallicity and elemental ratios of hydrogen-dominated envelopes. The third Tier of Ariel will be devoted to studying the best targets for atmospheric characterisation multiple times in search of temporal variations in chemistry or cloud coverage. Finally, a fourth tier will provide time to undertake observations which don't neatly fit into the original three tier process. Examples could include phase-curves or detailed studies of smaller planets which may host secondary atmospheres.

\begin{figure}
    \centering
    \includegraphics[width=0.45\textwidth]{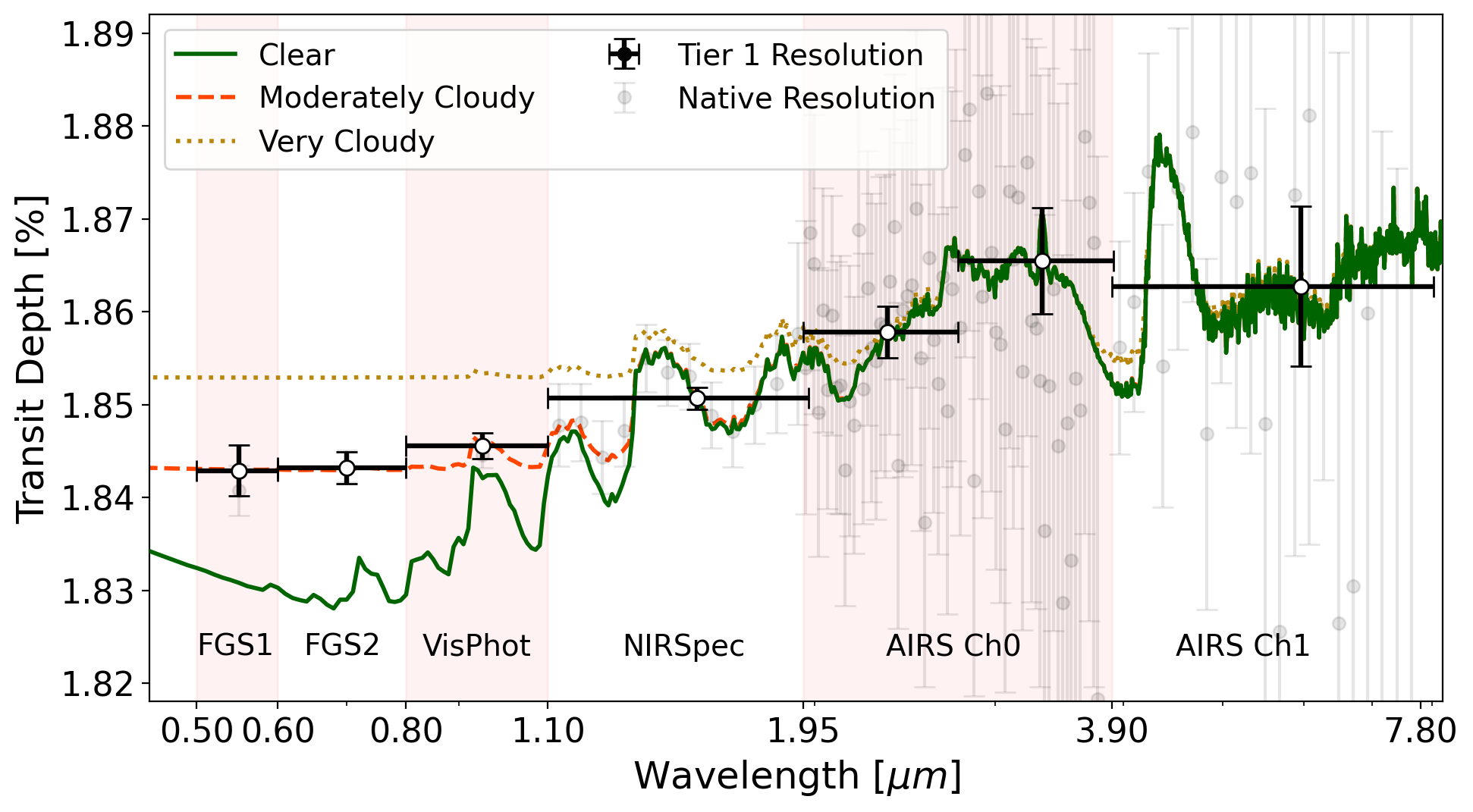}
    \includegraphics[width=0.45\textwidth]{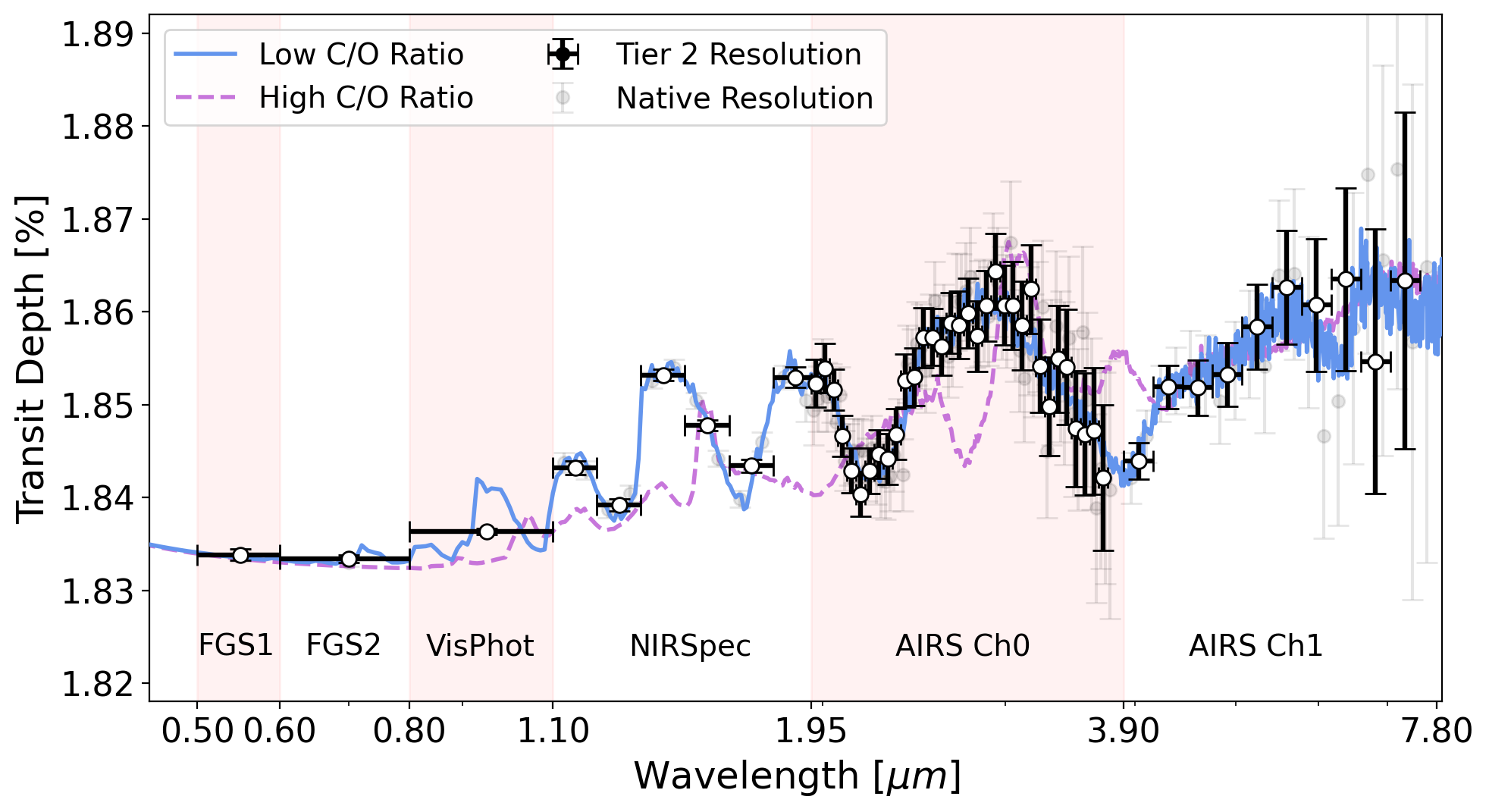}
    \caption{Examples of the spectral binning used to define the science requirements for Tier 1 (top) and 2 (bottom) of the Ariel survey. The reduction in resolution for these Tiers is a result of post-processing and so, for reference, the native spectral resolution at which all observations are taken is also shown. The error bars were computed using ArielRad and the atmospheric forward models computed using TauREx 3 \citep{al-refaie_taurex3}. The top panel is a free chemistry model while the TauREx GGChem plugin has been used for the bottom panel \citep{woitke_ggchem,al-refaie_taurex3_chem}. Random Gaussian scatter has been added to each spectra based on the magnitude of the spectral uncertainties. Ariel's instruments operate simultaneously and, therefore, the complete wavelength coverage (0.5-7.8 $\mu$m) is provided in a single observation.}
    \label{fig:t1_t2_data}
\end{figure}

The suitability of a planet for study with Ariel in each of these tiers is defined by a set of science requirements. While the requirements for Tier 3 are set at the native resolution of Ariel's instruments, Tiers 1 and 2 are set on data with a reduced resolution. The data quality requirement for each Tier is that, at the defined resolution, the expected SNR on the atmosphere of the planet is greater than 7. Examples of the data for Tier 1 and 2 are shown in Figure \ref{fig:t1_t2_data}. While the ESA science requirements are based upon the resolutions show in Figure \ref{fig:t1_t2_data}, the datasets could be analysed with different spectral binning if required as the reduction in resolution is accomplished during post-processing. More detail on the observation strategy of Ariel can be found in a number of studies \citep[e.g.][]{tinetti_ariel,tinetti_ariel2}.

In this work, the performance of the Ariel mission has been modelled using ArielRad \citep{mugnai_ar}. ArielRad is an adapted version of the instrument independent radiometric simulator ExoRad\footnote{https://github.com/ExObsSim/ExoRad2-public} and accounts for a wide variety of noise sources including those arising from the detector (readout, gain, dark current), the photon noise from the target star,  zodiacal background, instrument emission, and jitter noise. For each target, the expected uncertainty on as single transit or eclipse observation is determined at the native resolution of the instruments. It is assumed that Ariel observes for 2.5 times the length of the transit or eclipse. These uncertainties are then utilised to calculate the required number of observations to meet the SNR$>$7 requirement in each tier. The post-processing steps include reducing the resolution to those used for the Tier 1 and 2 requirements, generating error bars which can be used for more detailed analyses (e.g. spectral retrievals).

\subsection{The Mission Candidate Sample}

Having derived a list of targets which have all the parameters necessary for an ArielRad simulation, we use this efficient simulator to determine the number of observations required to reach the required signal-to-noise ratio in each tier. The boundary of suitability will not be concrete, with some planets being assessed on a case-by-case basis, but, to broadly understand the entire population, we set upper limits of 5 observations to achieve Tier 1 goals and 20 for Tier 2. These limits were also used in \citetalias{edwards_ariel_tl}. We use these requirements to construct the Mission Candidate Sample (MCS), a list of all potential targets for Ariel. Any planets which meet these requirements are henceforth considered suitable for observation with Ariel.

\begin{figure}
    \centering
    \includegraphics[width=0.45\textwidth]{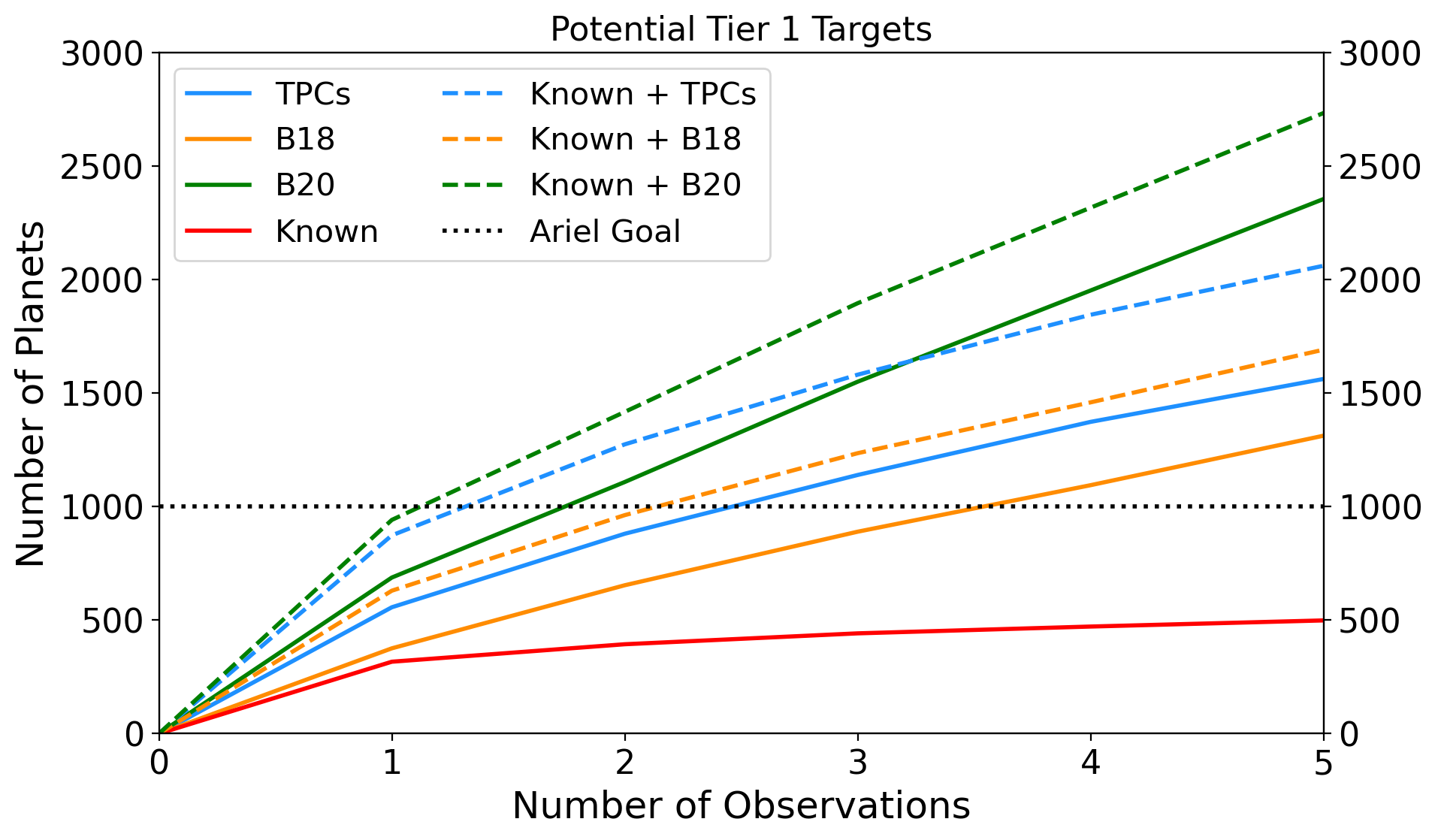}
    \includegraphics[width=0.45\textwidth]{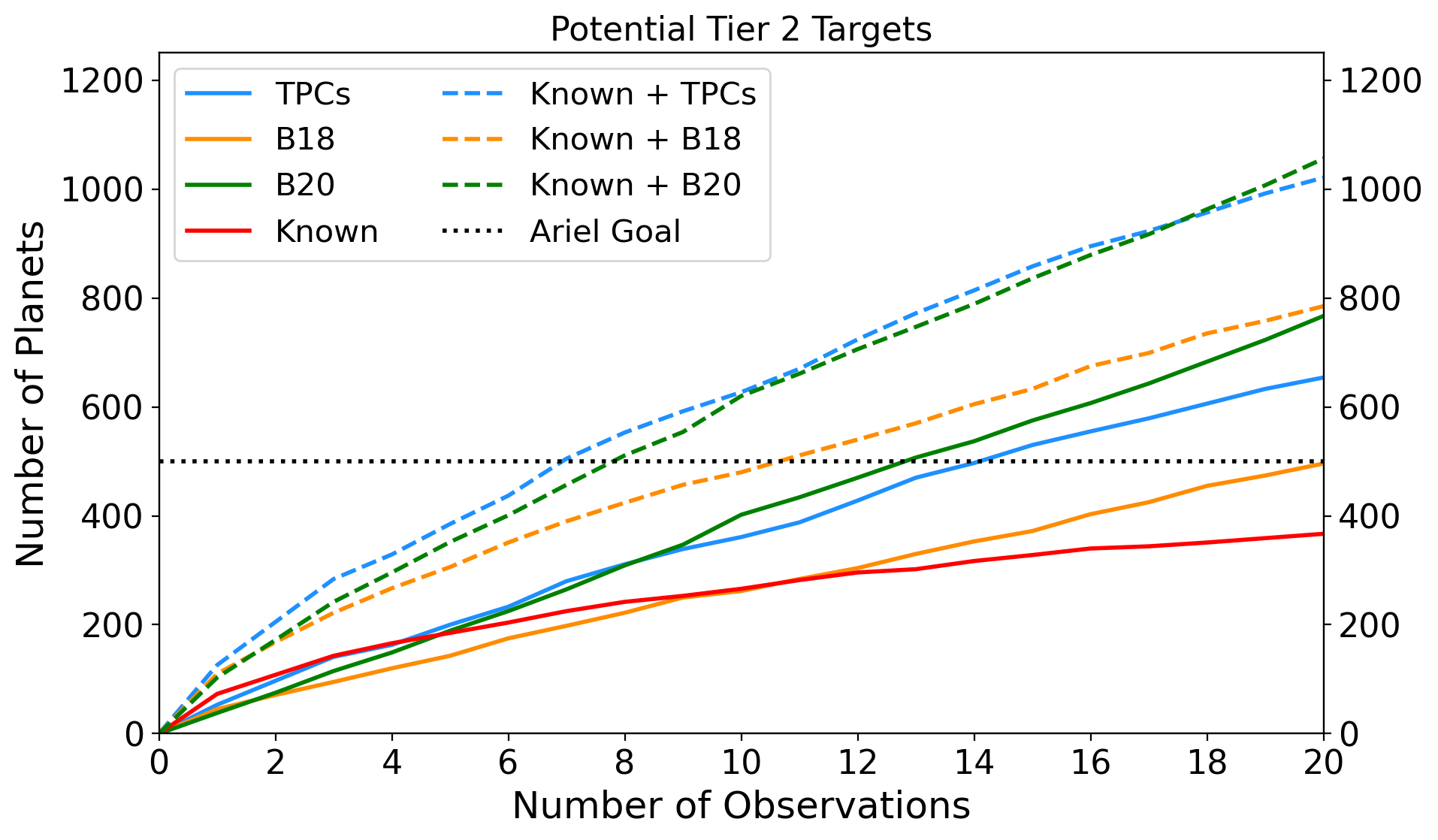}
    \caption{Planets suitable for study in Tier 1 (top) and Tier 2 (bottom). For lines marked by ``Known + B18" and ``Known + B20", we do not include any confirmed detections made by TESS. ``Known" and ``Known + TPCs" both include confirmed planets from the TESS mission.}
    \label{fig:t1_t2_planets}
\end{figure}

We compare the numbers of potential Tier 1 and 2 targets from \citetalias{barclay_2018} and \citetalias{barclay_2020} to the TPCs in Figure \ref{fig:t1_t2_planets}, showing the known population as well. From this we see that the number of TPCs that are suitable for study with Ariel currently lies between with the predictions of \citetalias{barclay_2018} and \citetalias{barclay_2020}. The situation is as one would expect given that TESS has finished its prime mission but has not yet completed its first extended mission. Analysis of 2241 TPCs from the prime mission by \citet{guerrero_tois} suggested the yield of targets which were suitable for atmospheric characterisation had not been as high as expected and our results suggest similar finding but that the extended mission has increased the number beyond the original yield prediction. However, we note that many TPCs may yet turn out to be false positives and the assumptions we have taken, particularly on parameters such as the mass, will also affect the final number of suitable targets.



\begin{figure}
    \centering
    \vspace{0.5cm}
    \includegraphics[width=0.45\textwidth]{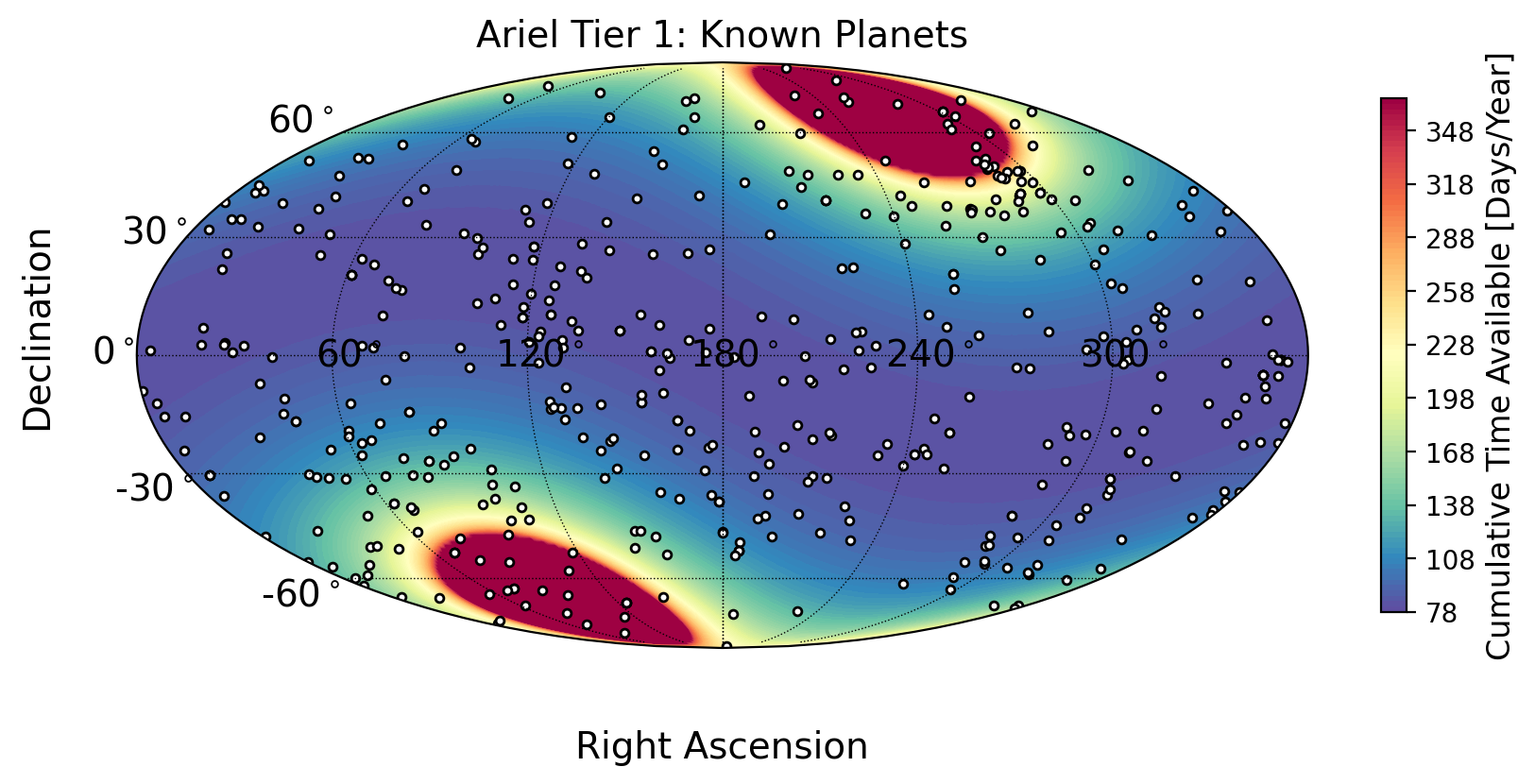}
    \includegraphics[width=0.45\textwidth]{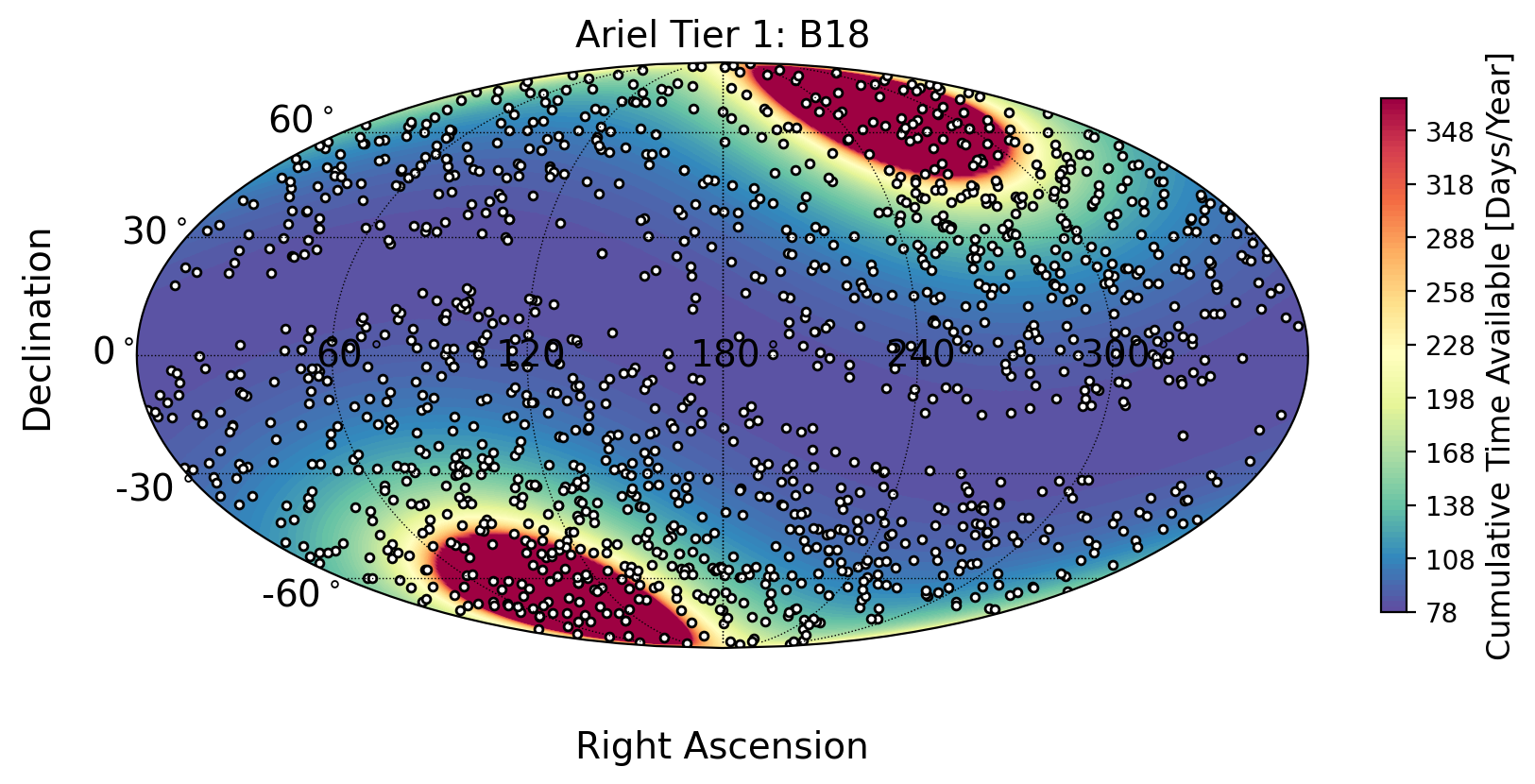}
    \includegraphics[width=0.45\textwidth]{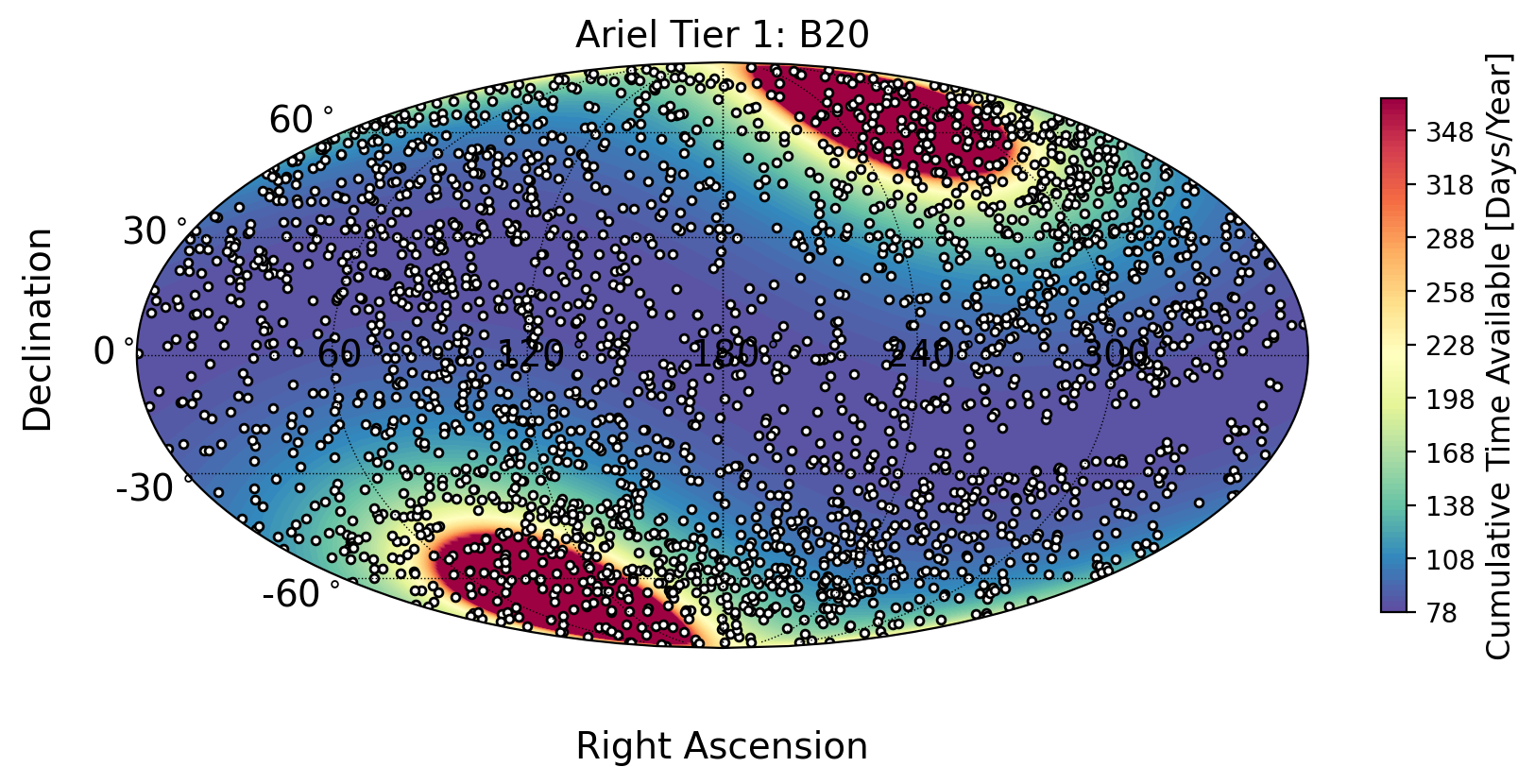}
    \includegraphics[width=0.45\textwidth]{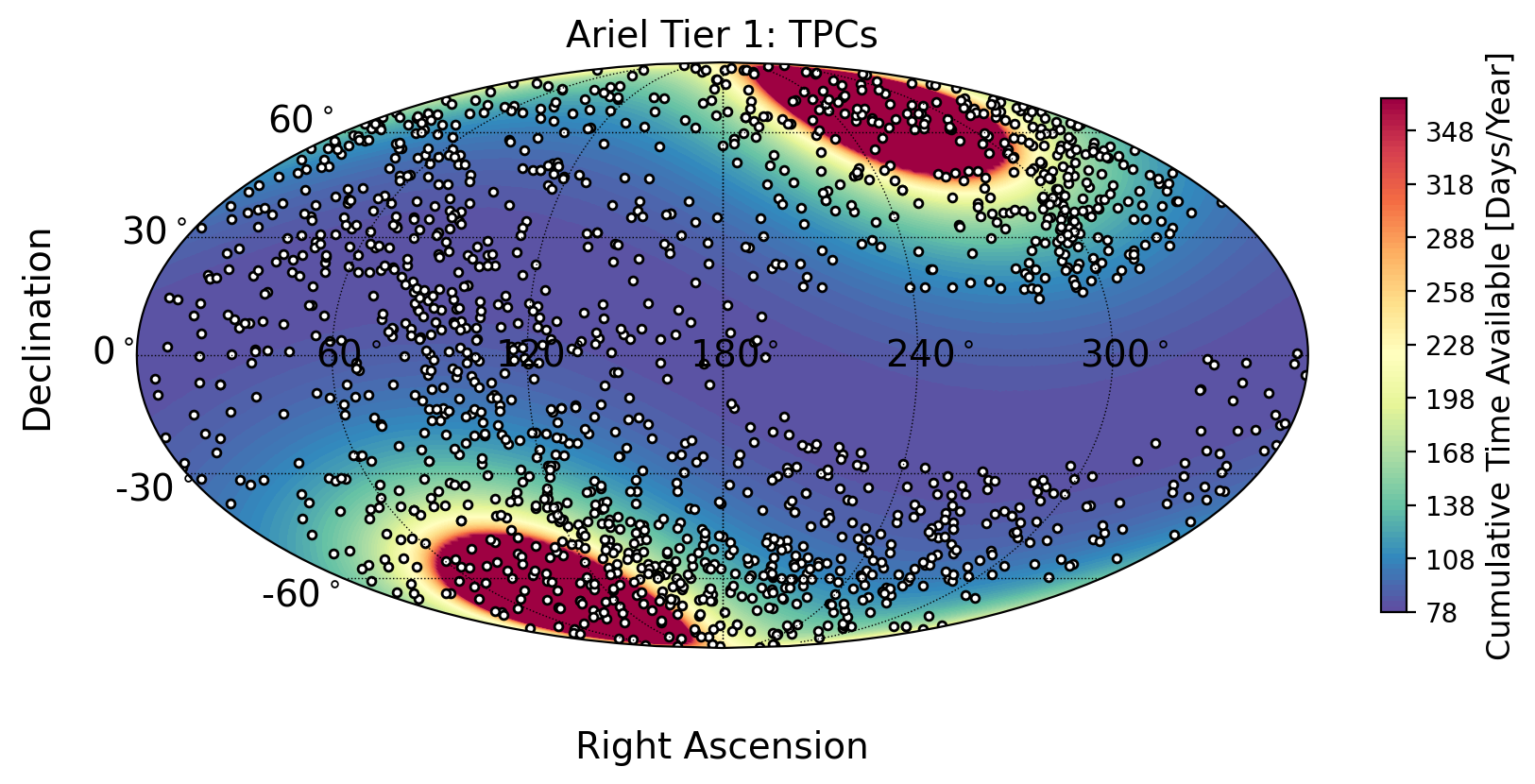}
    \caption{Sky locations of potential Ariel Tier 1 targets. Top panel: currently-known planets,  second panel: Tier 1 targets from the predicted targets of \citetalias{barclay_2018}, third panel: Tier 1 targets from the predicted targets of \citetalias{barclay_2020}, bottom: potential Tier 1 targets from the current TPCs. The sky coverage of Ariel, which offers large continuous viewing zones at the ecliptic poles, was determined using the Terminus code \citep{edwards_terminus}. Ariel's continuous viewing zones have been heavily studied by the TESS mission, leading to many candidates within them.}
    \label{fig:t1_sky_loc}
\end{figure}

The large number of TPCs is impressive given the effects stray light from the Earth and Moon have had on TESS. Excessive contamination by these sources caused Sectors 14-16 and 24-26 to be shifted northwards by around 30$^\circ$, leaving a portion of the sky unobserved during the prime mission \citep{guerrero_tois}. The gap this shift caused can be seen when comparing the sky locations of the Tier 1 targets, which is shown in Figure \ref{fig:t1_sky_loc}. Nevertheless, during the first three years of operation TESS has undoubtedly provided an enormous number of planet candidates which could be suitable for atmospheric characterisation. Additionally, the extended mission is covering the portions of the sky missed previously due to stray light as well as the ecliptic plane, leading to further detections, some of which may be suitable for study with Ariel.

\begin{figure}
    \centering
    \vspace{5mm}
    \includegraphics[width=0.45\textwidth]{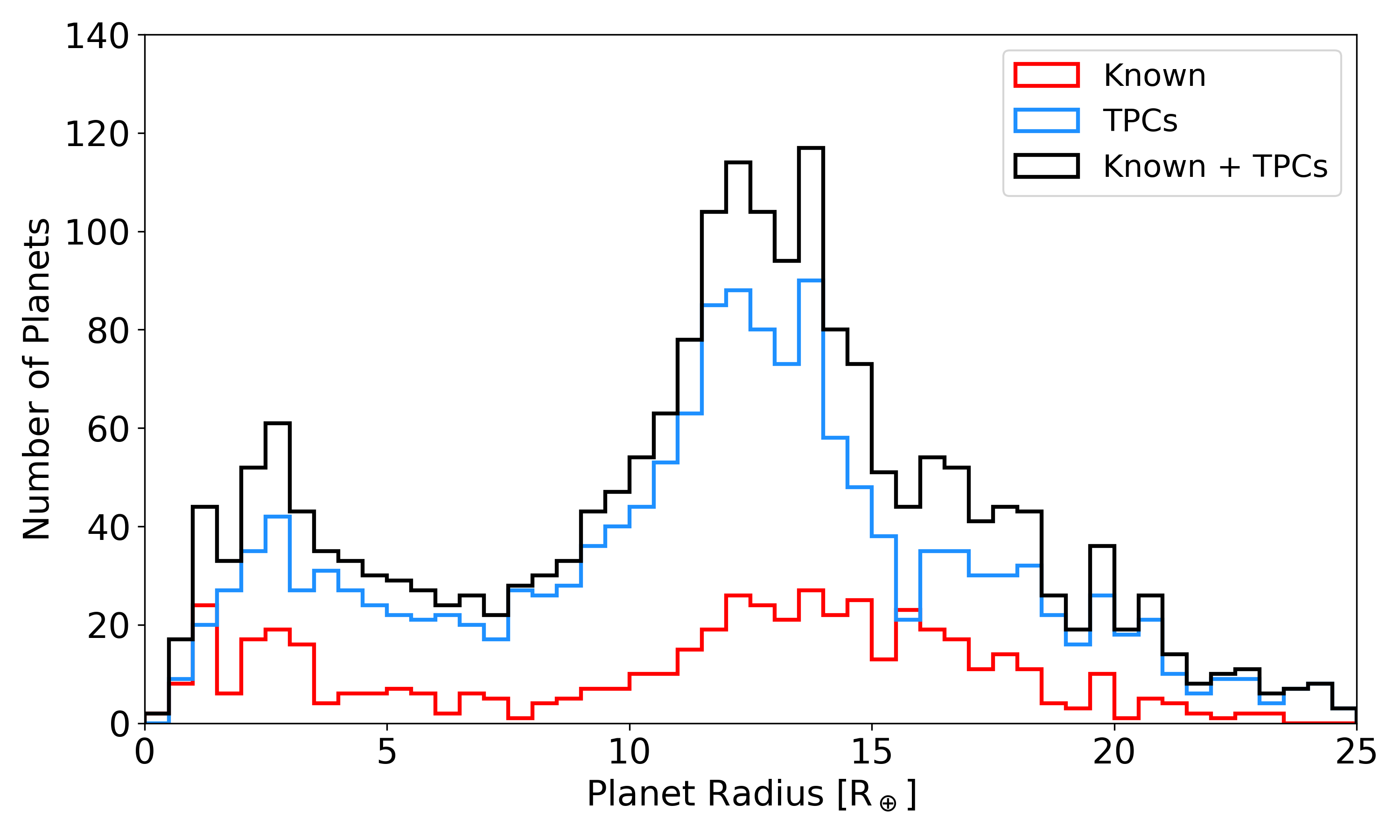}
    \includegraphics[width=0.45\textwidth]{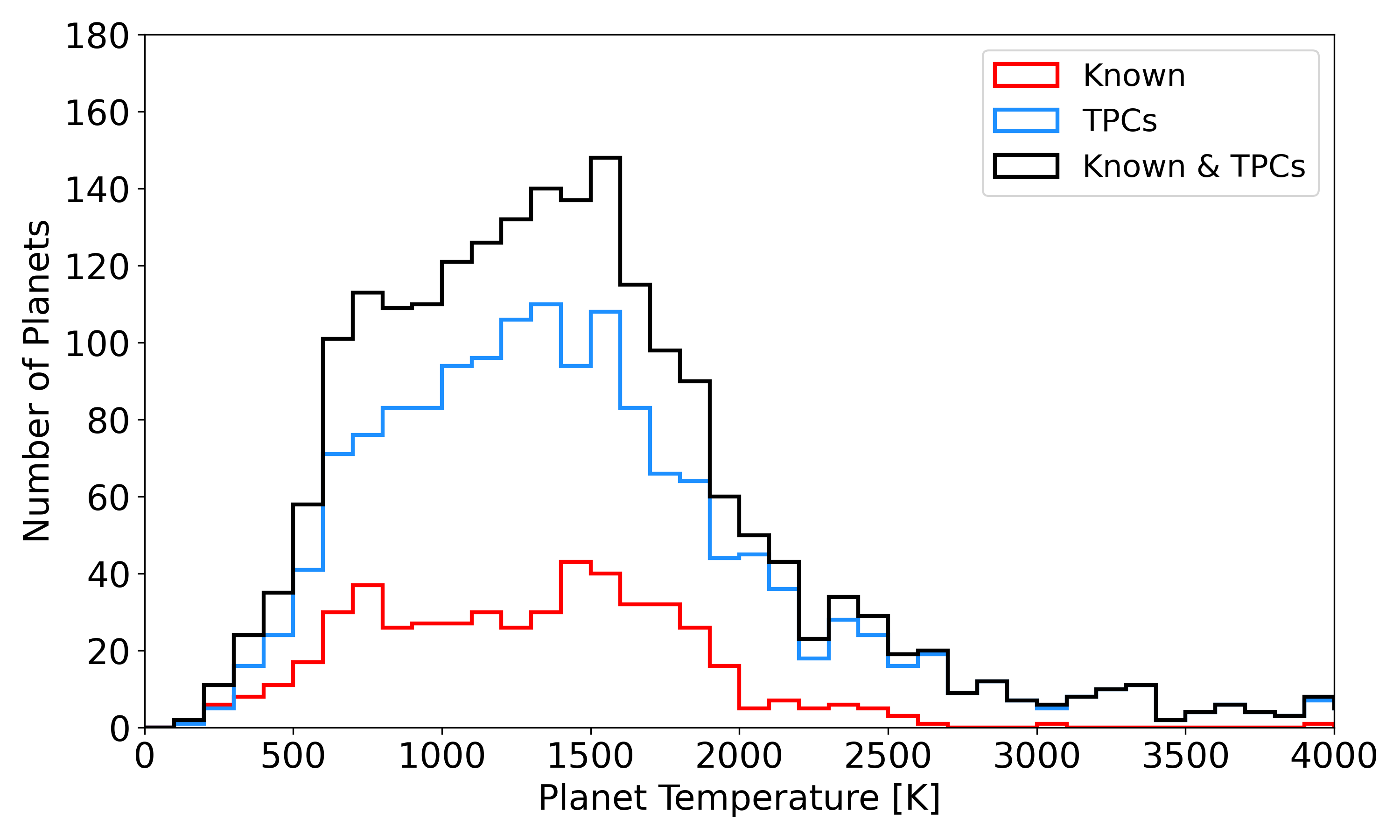}
    \includegraphics[width=0.45\textwidth]{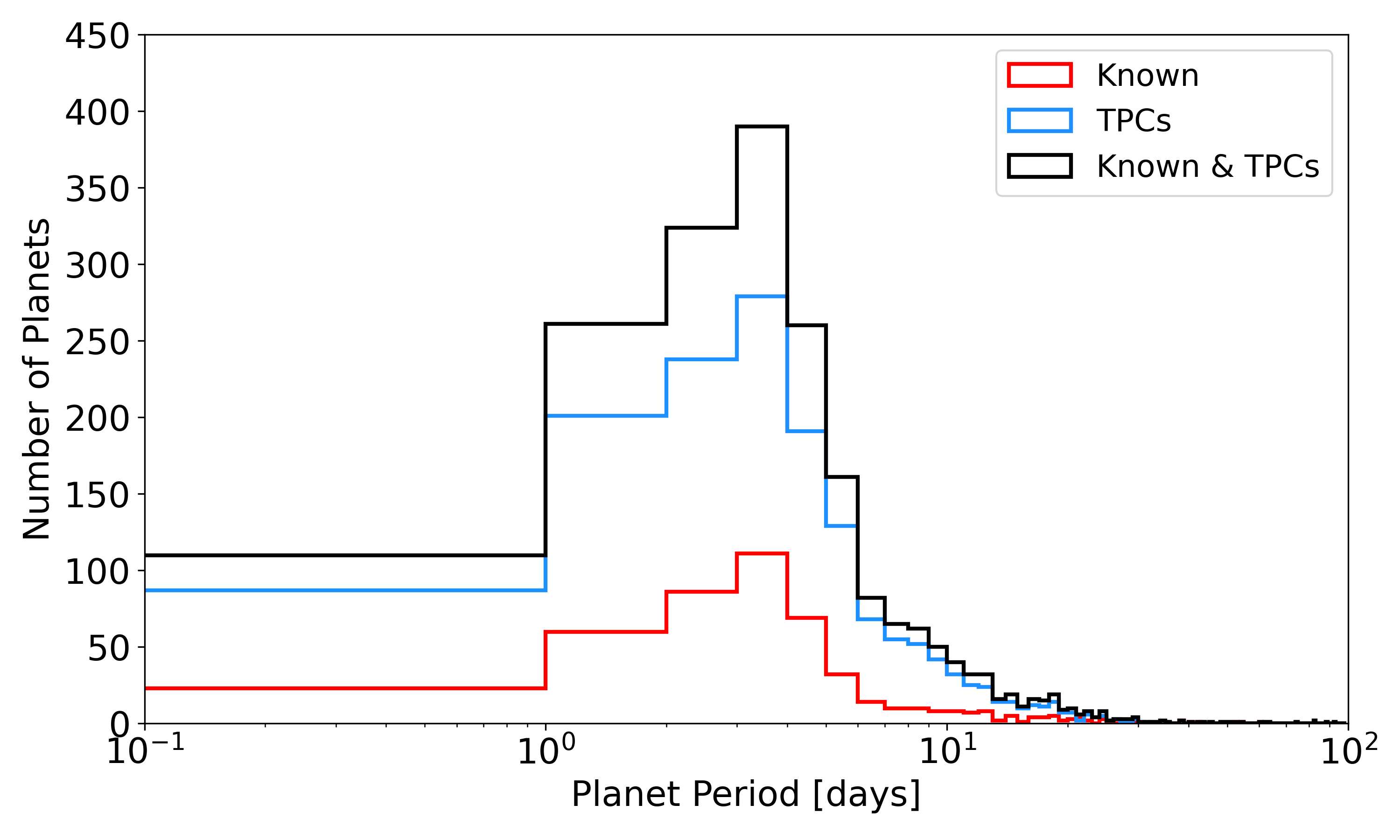}
    \caption{Histograms of the properties of the planets within the potential Ariel Tier 1 Catalogue. While the period of planets that can be studied are general constraint to $<20$ days, in terms of planet temperature and radii the targets will be highly diverse.}
    \label{fig:t1_distribution_pl}
    \vspace{5mm}
\end{figure}

\begin{figure}
    \centering
    \vspace{5mm}
    \includegraphics[width=0.45\textwidth]{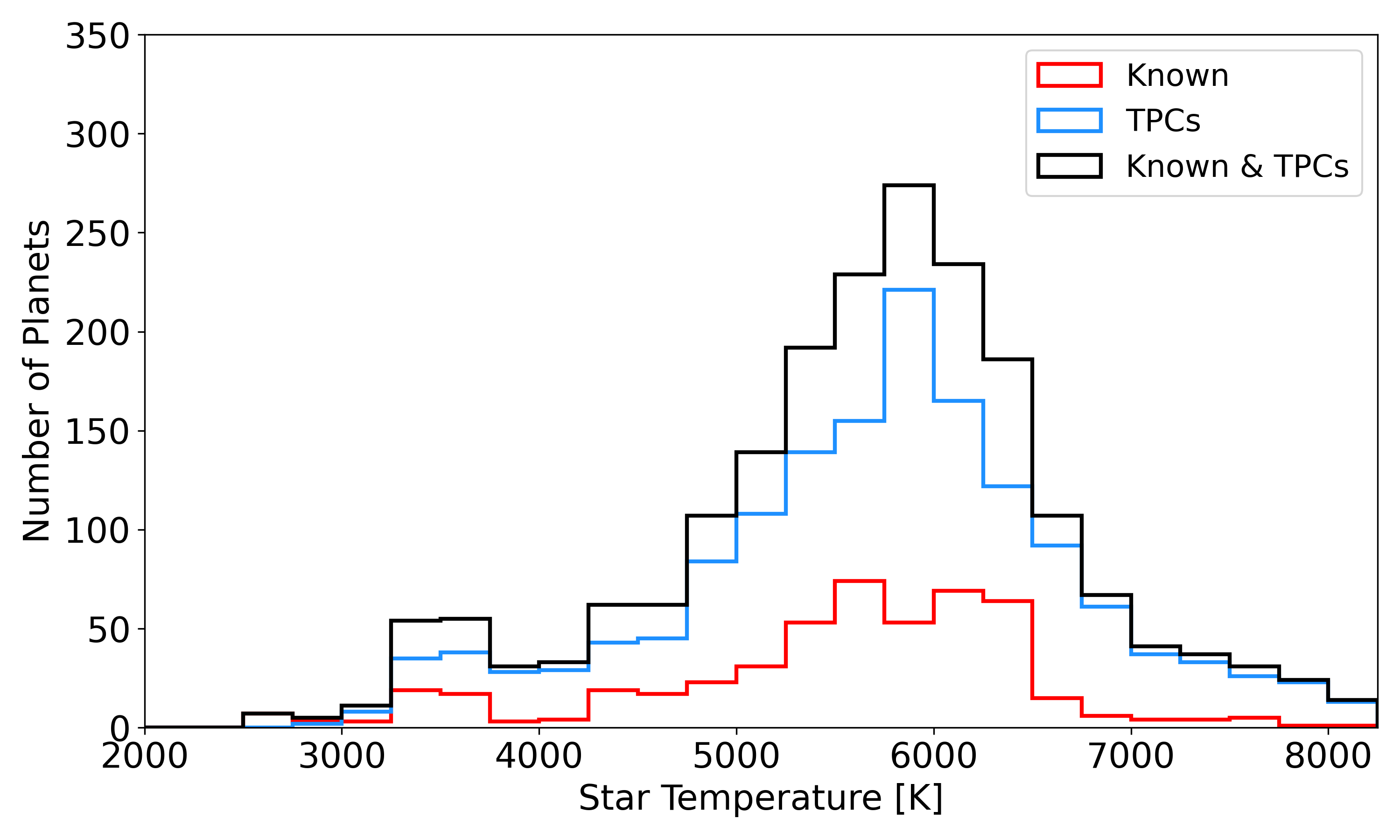}
    \includegraphics[width=0.45\textwidth]{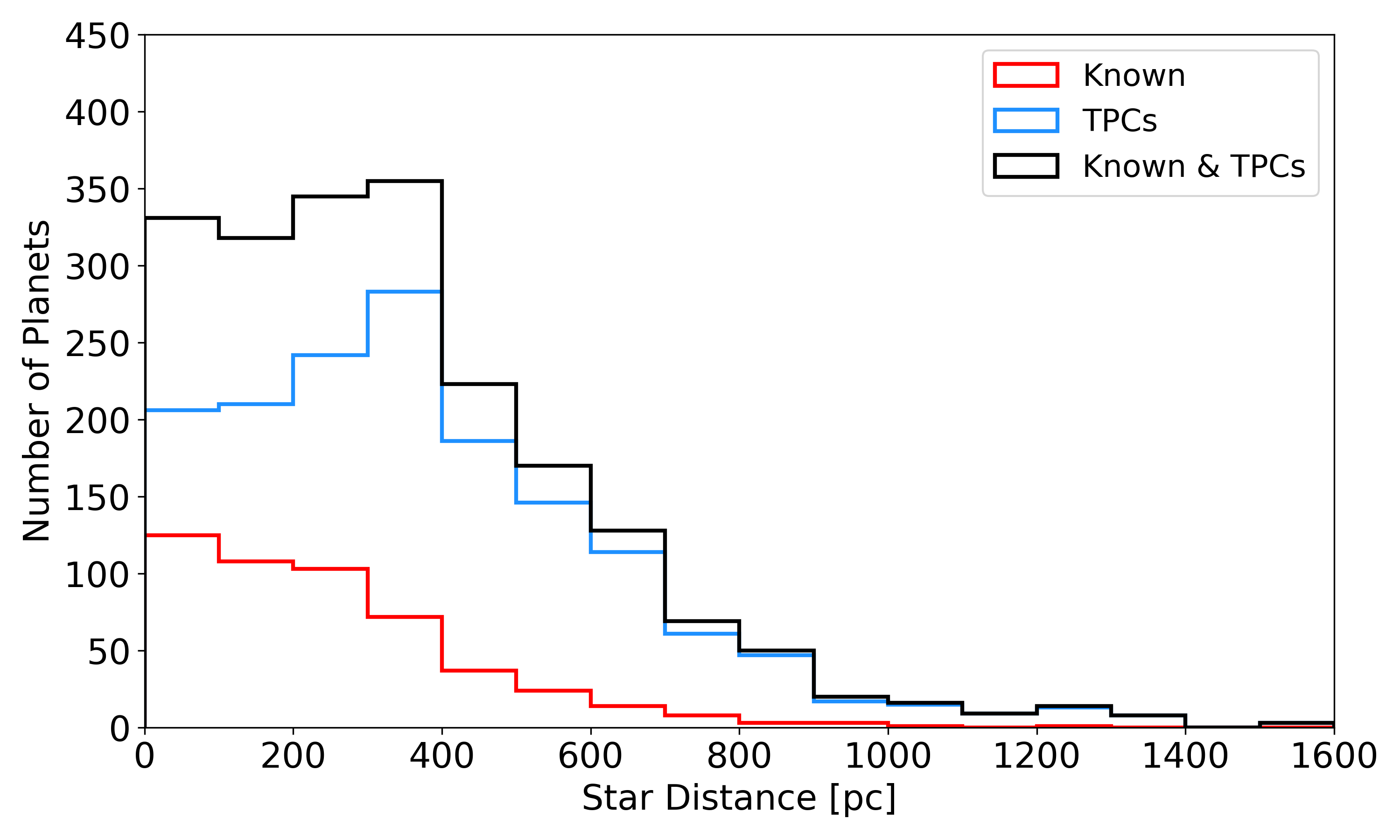}
    \includegraphics[width=0.45\textwidth]{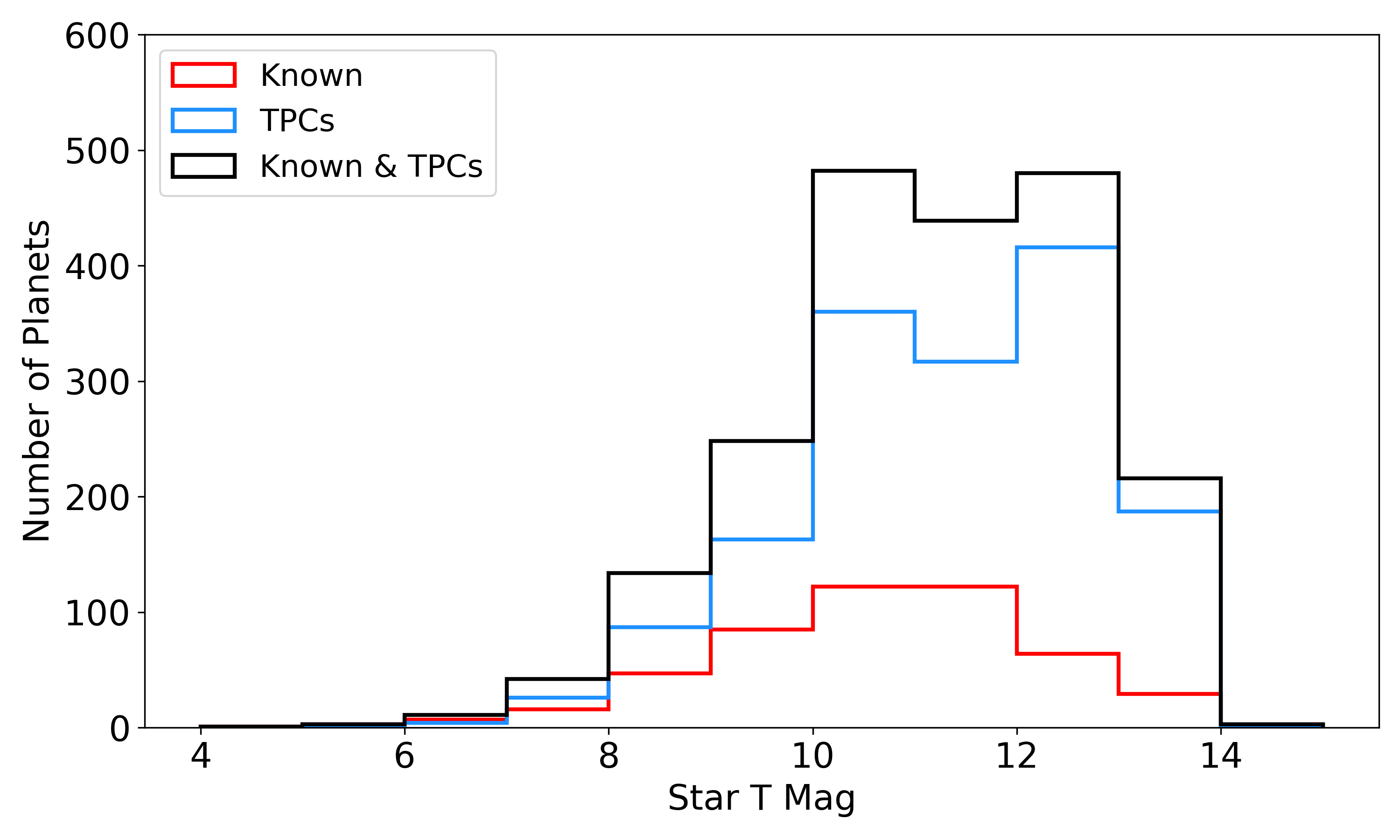}
    \caption{Histograms of the properties of the host stars within the potential Ariel Tier 1 Catalogue. Ariel will probe planets around a range of stellar types, out to around 1000\,pc from the Earth. The host star could be as faint as T$_{\rm mag}\sim$14.}
    \label{fig:t1_distribution_st}
    \vspace{5mm}
\end{figure}

Currently, around 500 confirmed planets would meet Ariel's Tier 1 requirements in 5 observations or fewer and, based on our assumptions, over 1700 TPCs could as well. Hence, even if half of these TPCs are false positives, it is likely there will be a copious number of planets to choose between for study with Ariel. Therefore, we now focus on the variety of targets that could be studied. The histograms in Figures \ref{fig:t1_distribution_pl} and \ref{fig:t1_distribution_st} highlight the range of Tier 1 targets that are available from the currently known population and the current TPCs. When combining these together, it is appears that a wide parameter space could be probed by Ariel. While the planets studied will generally have a short period ($<$ 20 days), their temperatures will vary from 200 to 4000 K and span radii from Earth-sized to super-inflated Jupiters. The stellar hosts primarily have temperature between 4000 K and 7000 K, but cooler dwarfs, as well as A-type stars, are also present. We note that some targets around cooler or fainter stars have been deemed unsuitable for study due to the brightness requirements of Ariel's FGS channels which facilitate pointing of the spacecraft. Whether these requirements were reached was assessed by using ArielRad and those that failed to reach them are not included in our analysis.



In Tier 1, Ariel will attempt to study planets in every category of a five dimensional parameter space. They will be classified by the planet's radius, density and temperature, as well as the host star's temperature and metallicity. However, as the TPCs do not have measured masses, and their stellar metallicities are not listed in the TPC catalogue, we have to overlook these parameters for now. Therefore, we classified the potential Tier 1 targets in a three dimensional parameter space using the classifications shown in Table \ref{tab:star_para}, with the bounds for the stellar types being set by using the values from \citet{pecaut_stars}\footnote{\url{http://www.pas.rochester.edu/~emamajek/EEM_dwarf_UBVIJHK_colors_Teff.txt}}. 

\begin{table*}[]
    \centering
    \begin{tabular}{cccccccc}\hline \hline
     Star Type & Temperature Bounds & & Planet Type & Radius Bounds & & Climate & Temperature Bounds\\\hline
     M & T$_{\rm s}$ $\leq$ 3890 K & & Earth/Super-Earth & R$_{\rm p}$ $<$ 1.8 R$_\oplus$ & & Warm & T$_{\rm P}$ $\leq$500 K\\
     K & 3890 K $<$ T$_{\rm s}$ $\leq$5330 K & & Sub-Neptune & 1.8 R$_\oplus$ $<$ R$_{\rm p}$ $\leq$ 3.5 R$_\oplus$ & & Very Warm & 500 K $<$ T$_{\rm P}$ $\leq$ 1000 K\\
     G & 5330 K $<$ T$_{\rm s}$ $\leq$ 5960 K & & Neptune & 3.5 R$_\oplus$ $<$ R$_{\rm p}$ $\leq$ 6 R$_\oplus$ & & Hot & 1000 K $<$ T$_{\rm P}\leq$ 1500 K\\
     F & 5960 K $<$ T$_{\rm s}\leq$7300 K & & Jupiter & 6 R$_\oplus$ $<$ R$_{\rm p}$ $\leq$ 16 R$_\oplus$ & & Very Hot & 1500 K $<$ T$_{\rm P}\leq$ 2500 K\\
     A & 7300 K $<$ T$_{\rm s}$ & & Inflated Jupiter & 16 R$_\oplus$ $<$ R$_{\rm p}$ $\leq$ 26 R$_\oplus$ & & Ultra Hot & 2500 K $<$ T$_{\rm P}\leq$ 4000 K\\\hline \hline
    \end{tabular}
    \caption{Bounds used to categorise host stars in Figures \ref{fig:distribution_mkg} and \ref{fig:distribution_fa}. }
    \label{tab:star_para}
\end{table*}

\begin{figure*}
    \centering
    \includegraphics[width=0.45\textwidth]{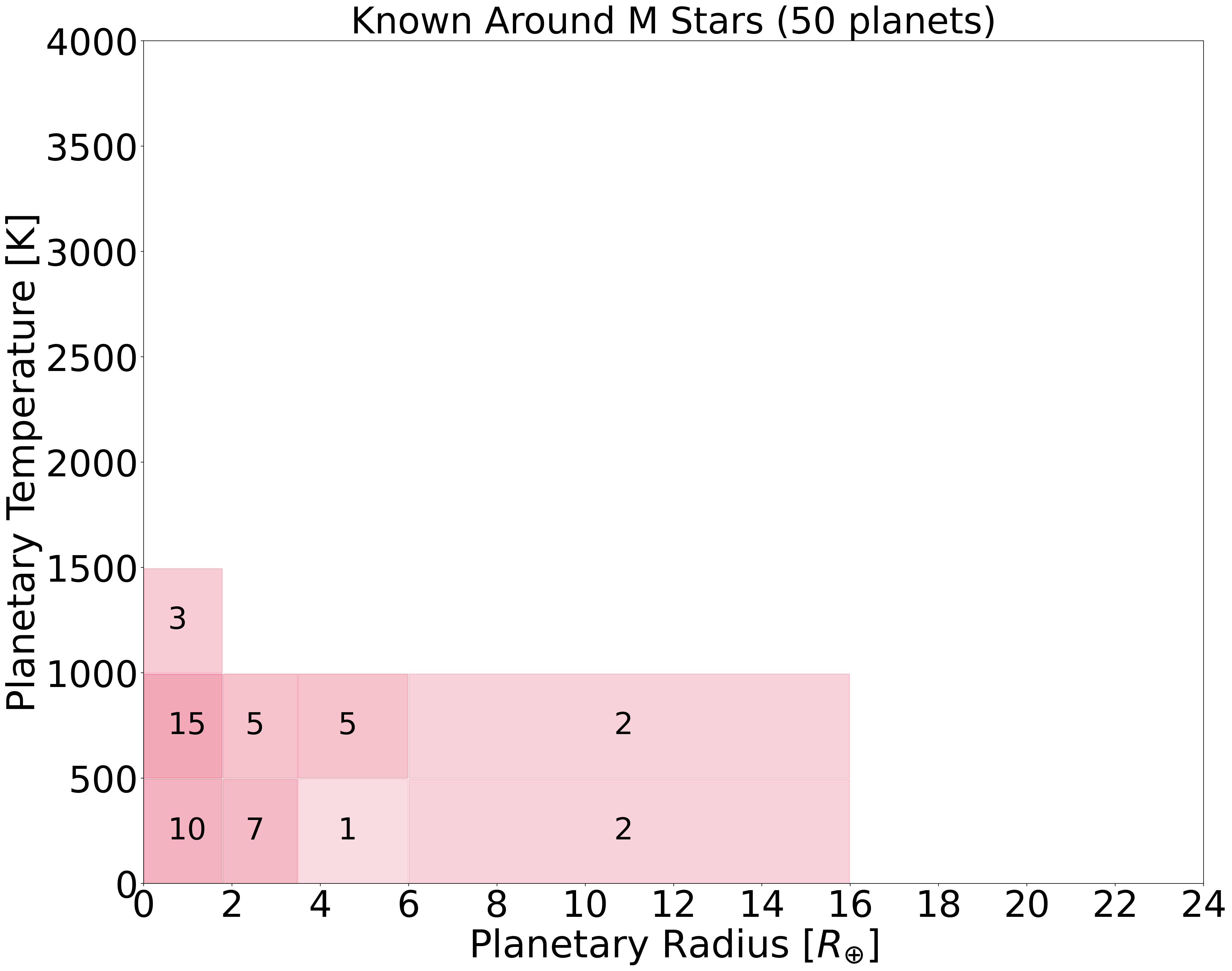}
    \includegraphics[width=0.45\textwidth]{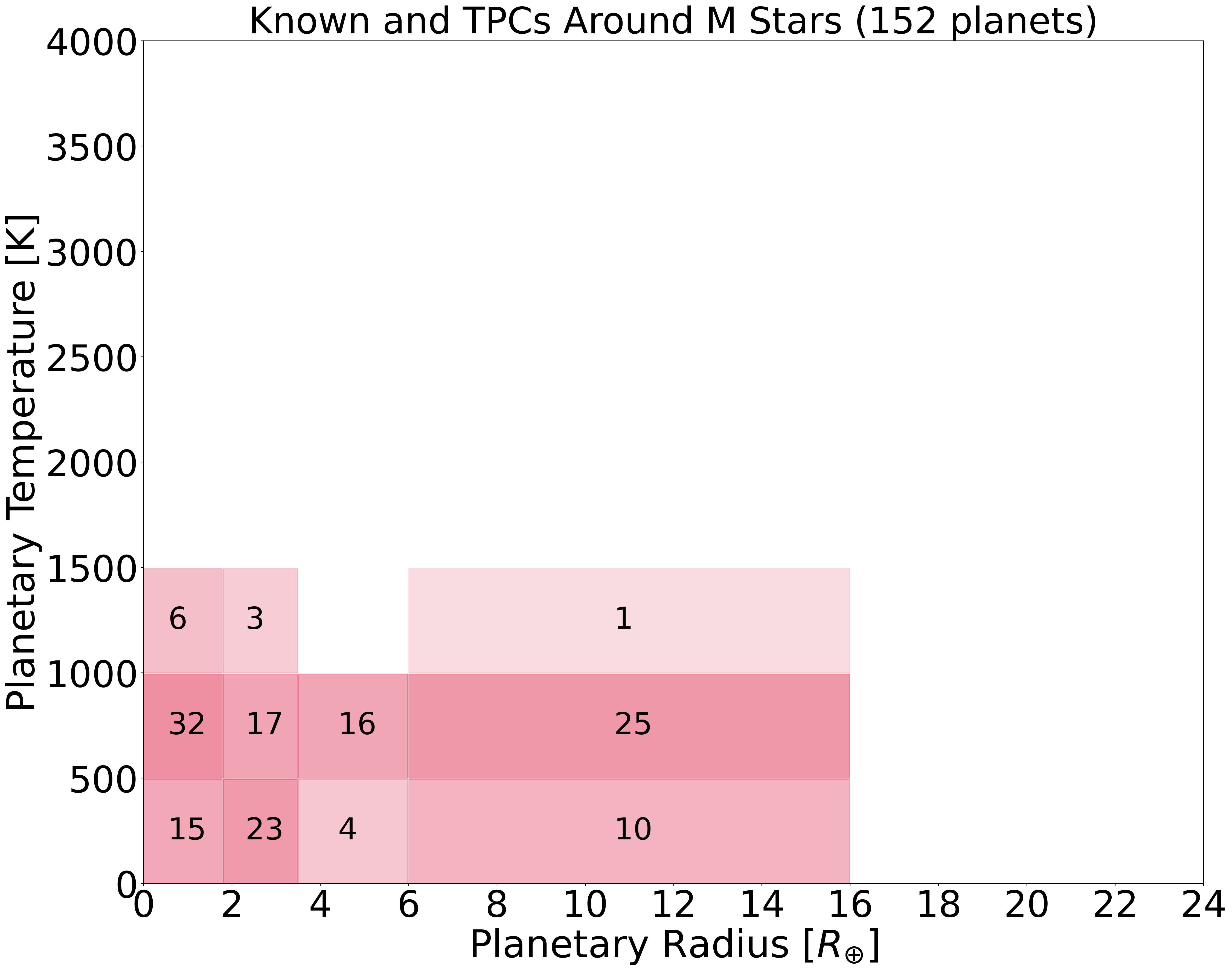}
    \includegraphics[width=0.45\textwidth]{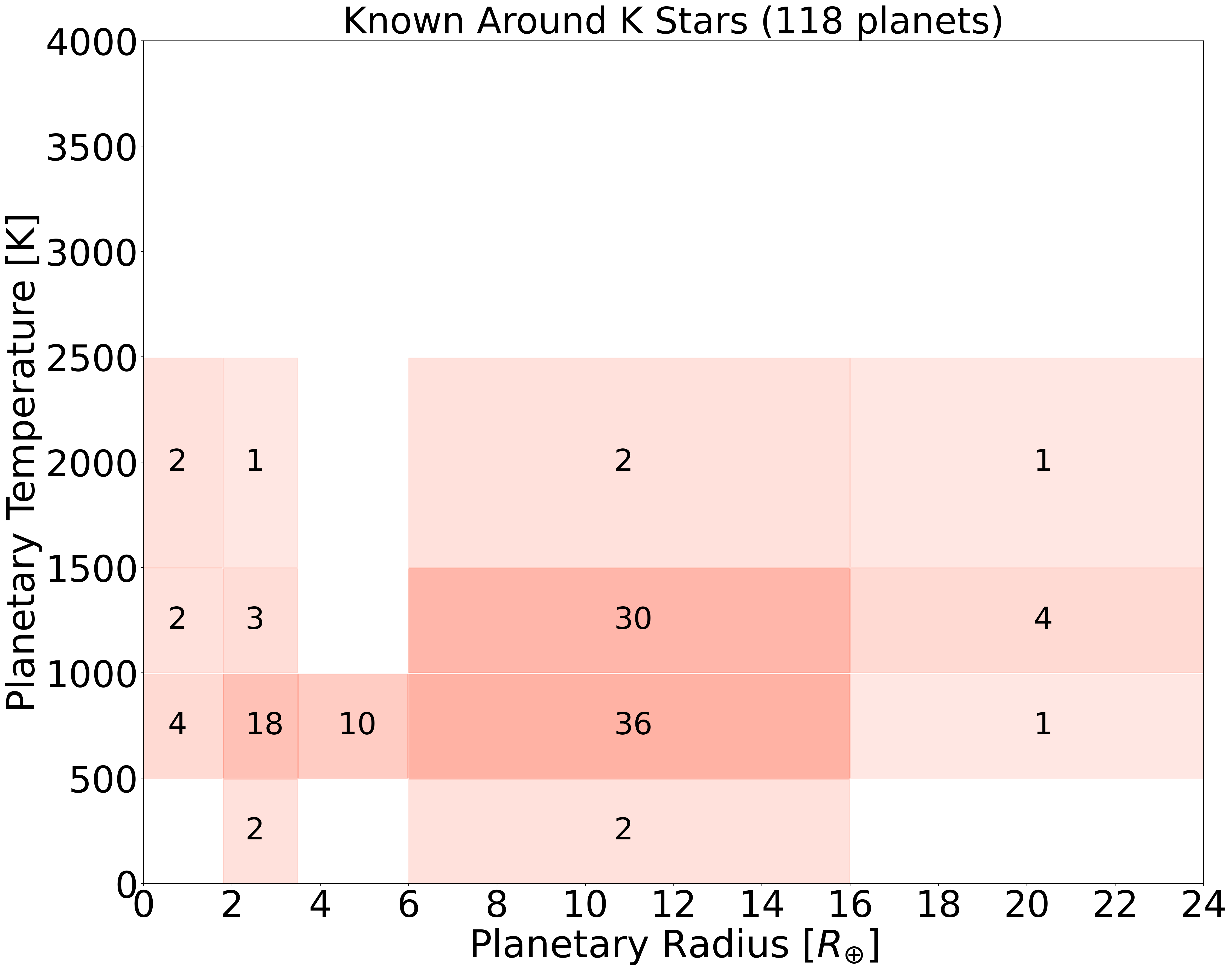}
    \includegraphics[width=0.45\textwidth]{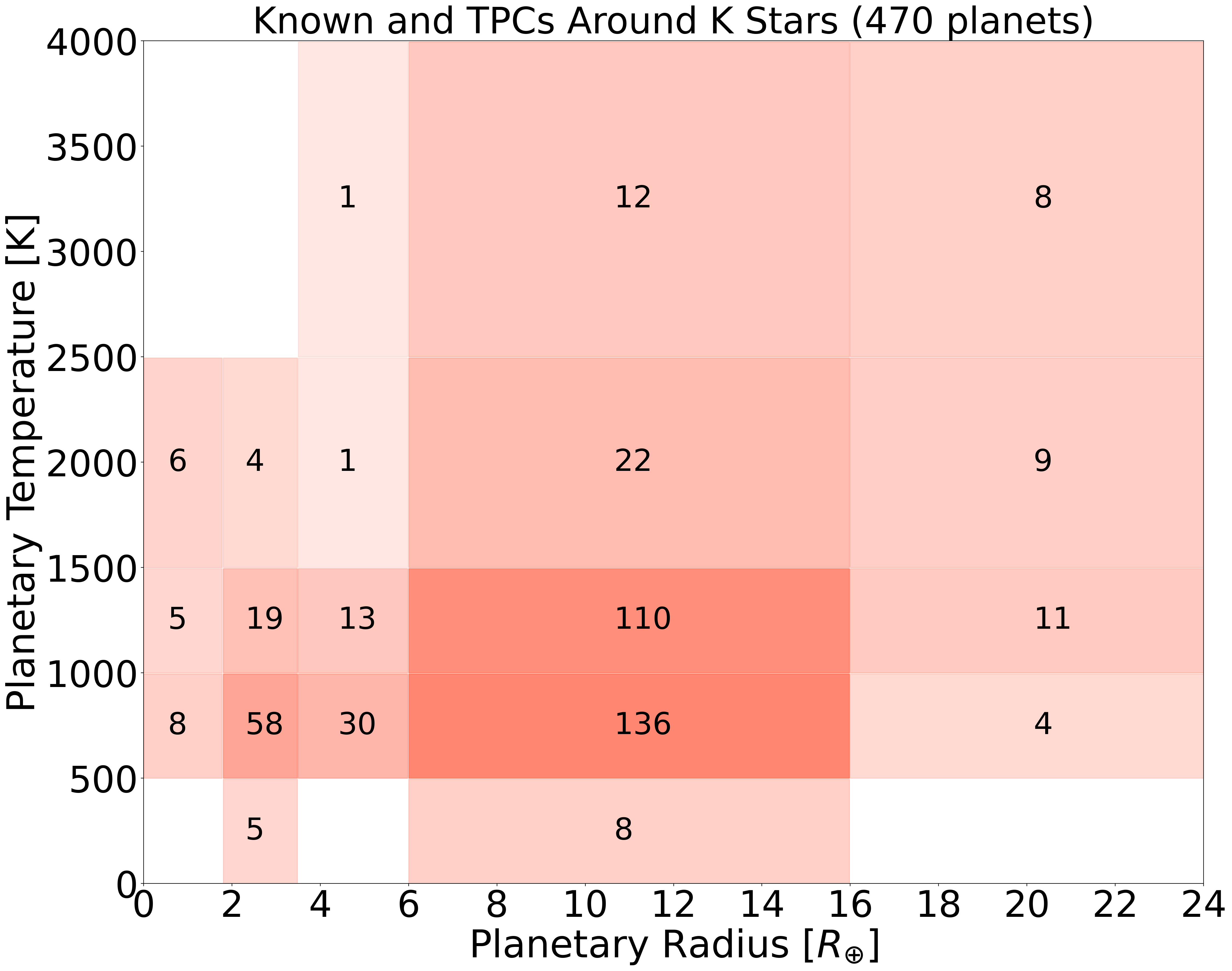}
    \includegraphics[width=0.45\textwidth]{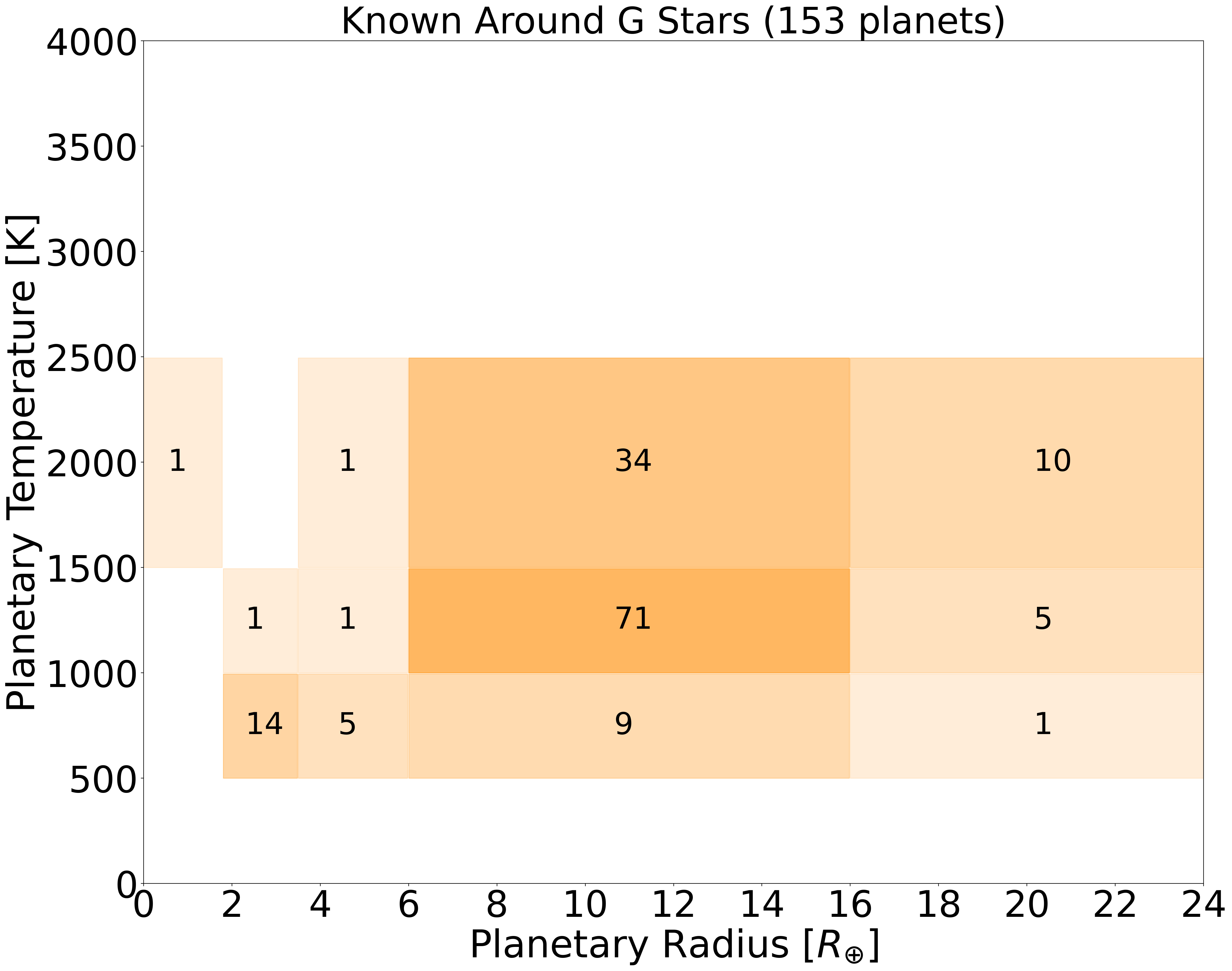}
    \includegraphics[width=0.45\textwidth]{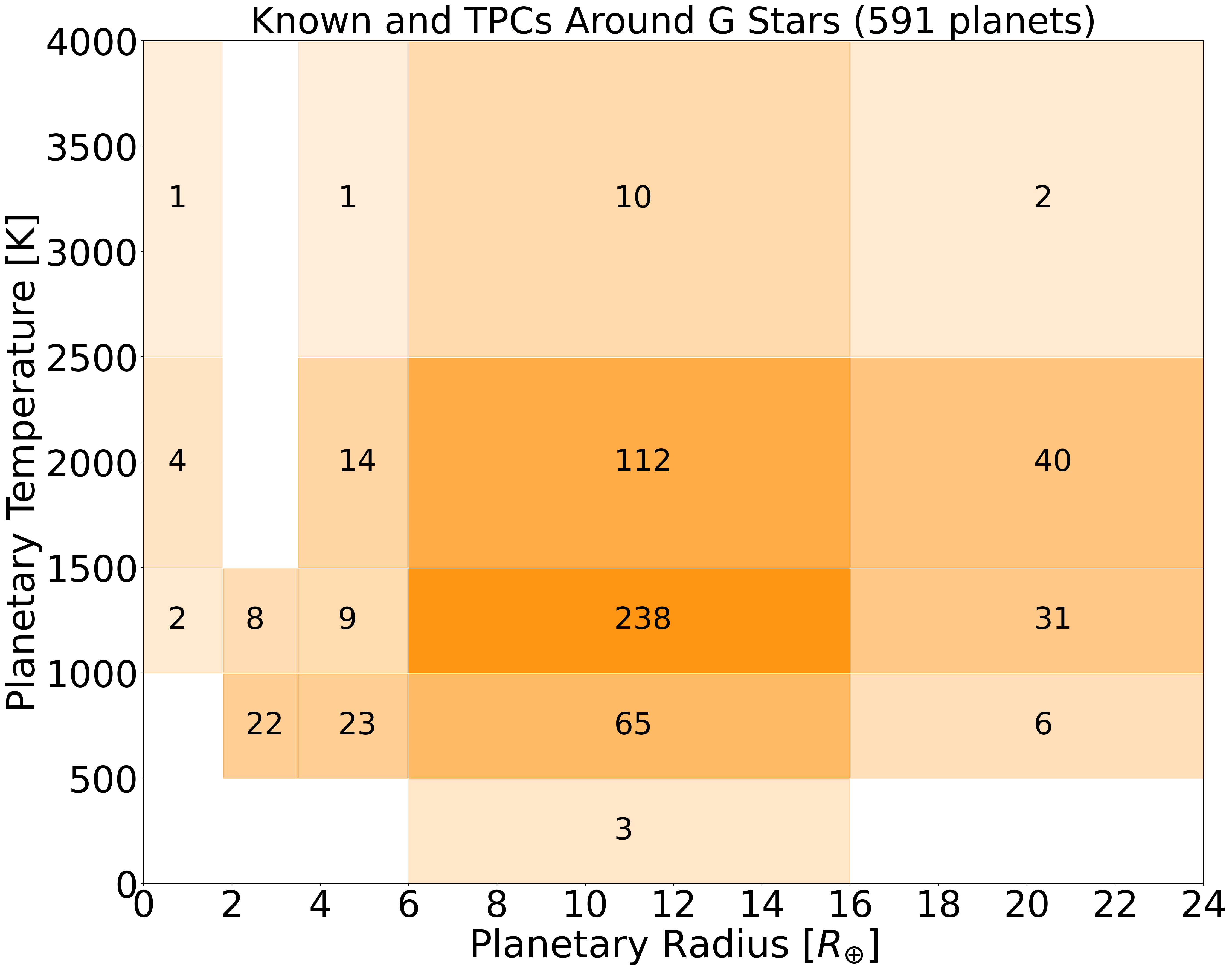}
    \caption{Temperature and radius distribution of known planets, including those found by TESS, (left) and these known planets in addition to the current TPCs (right), for M (top), K (middle) and G (bottom) stars, that are suitable for Tier 1 study with Ariel.}
    \label{fig:distribution_mkg}
\end{figure*}

\begin{figure*}
    \centering
    \includegraphics[width=0.45\textwidth]{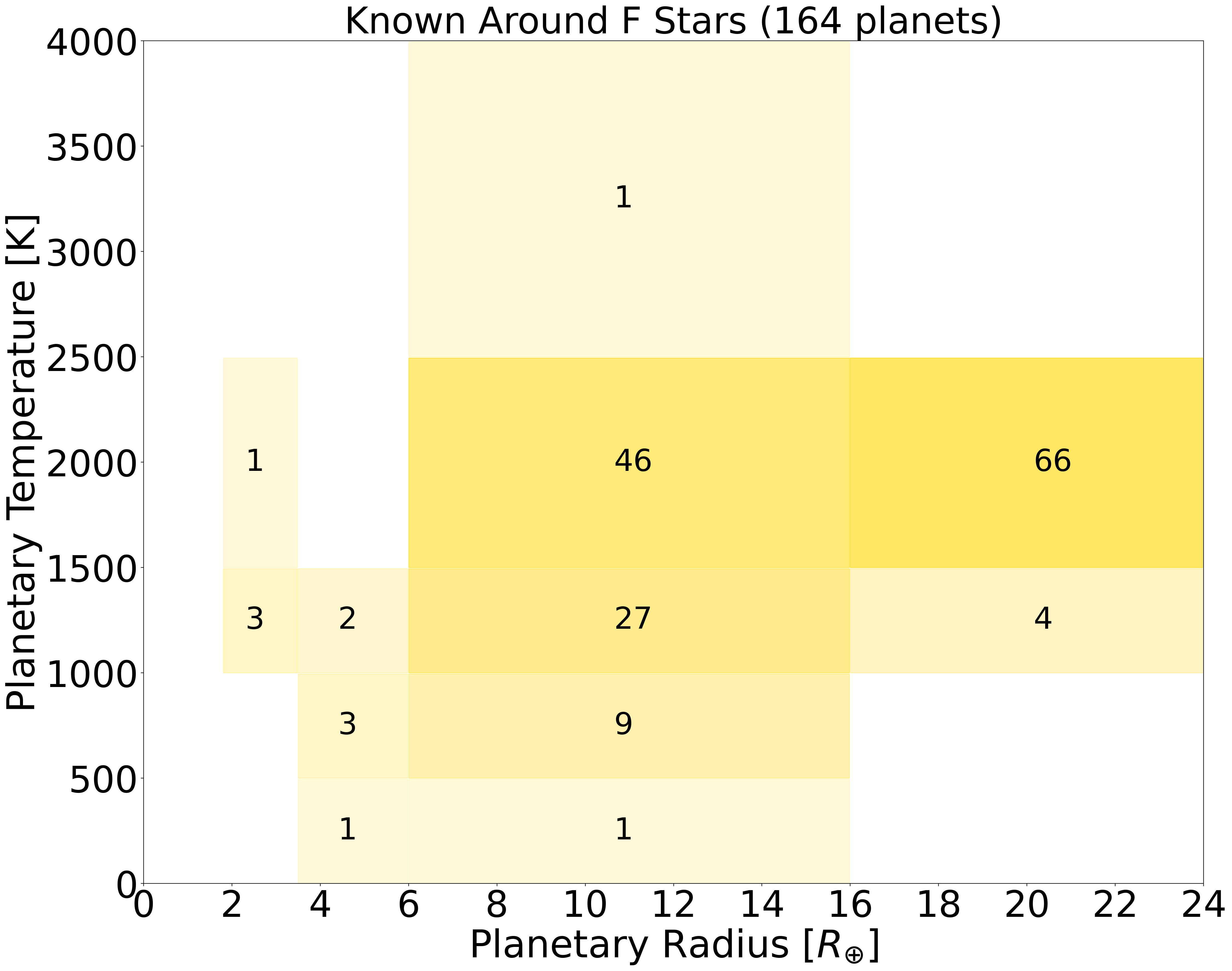}
    \includegraphics[width=0.45\textwidth]{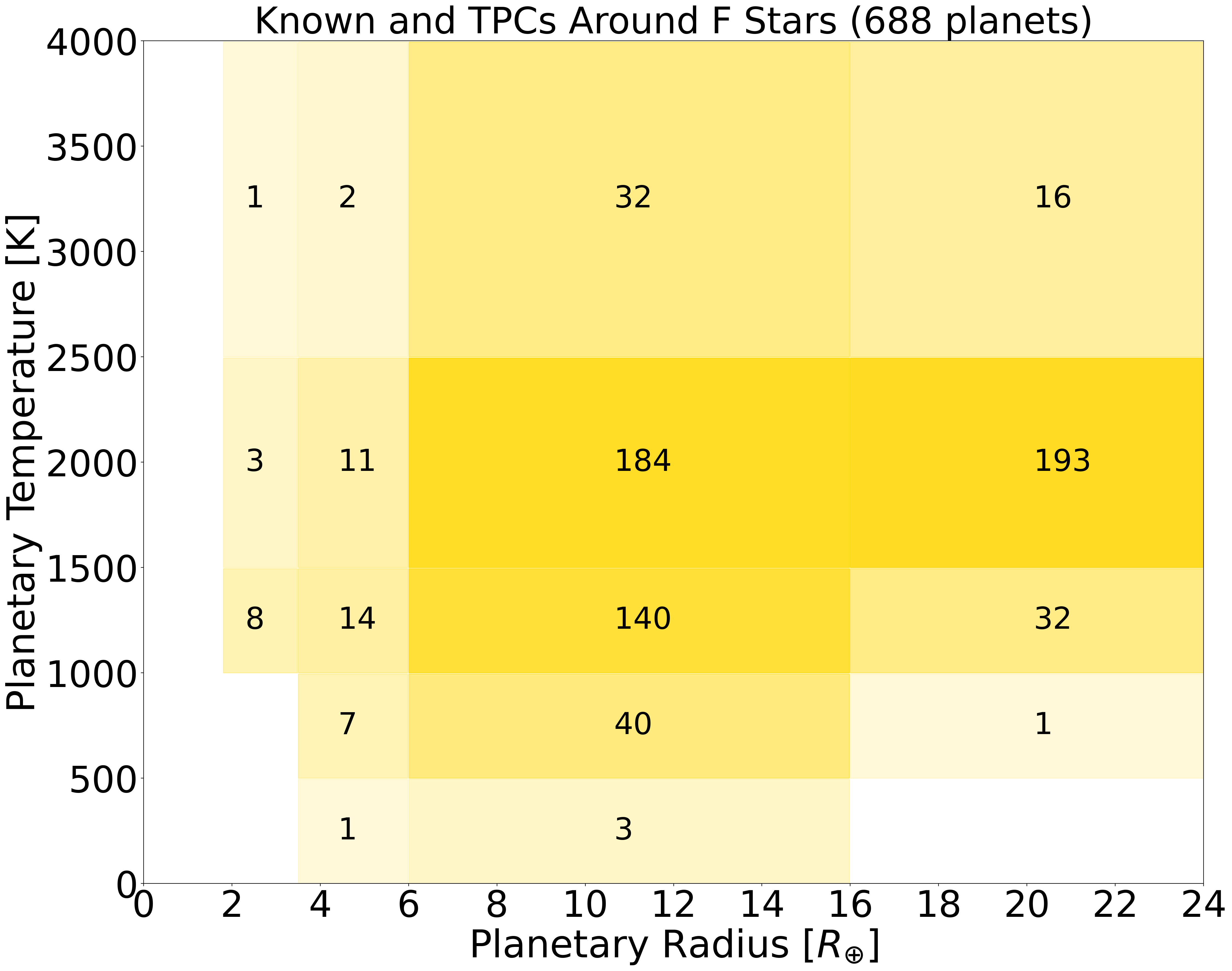}
    \includegraphics[width=0.45\textwidth]{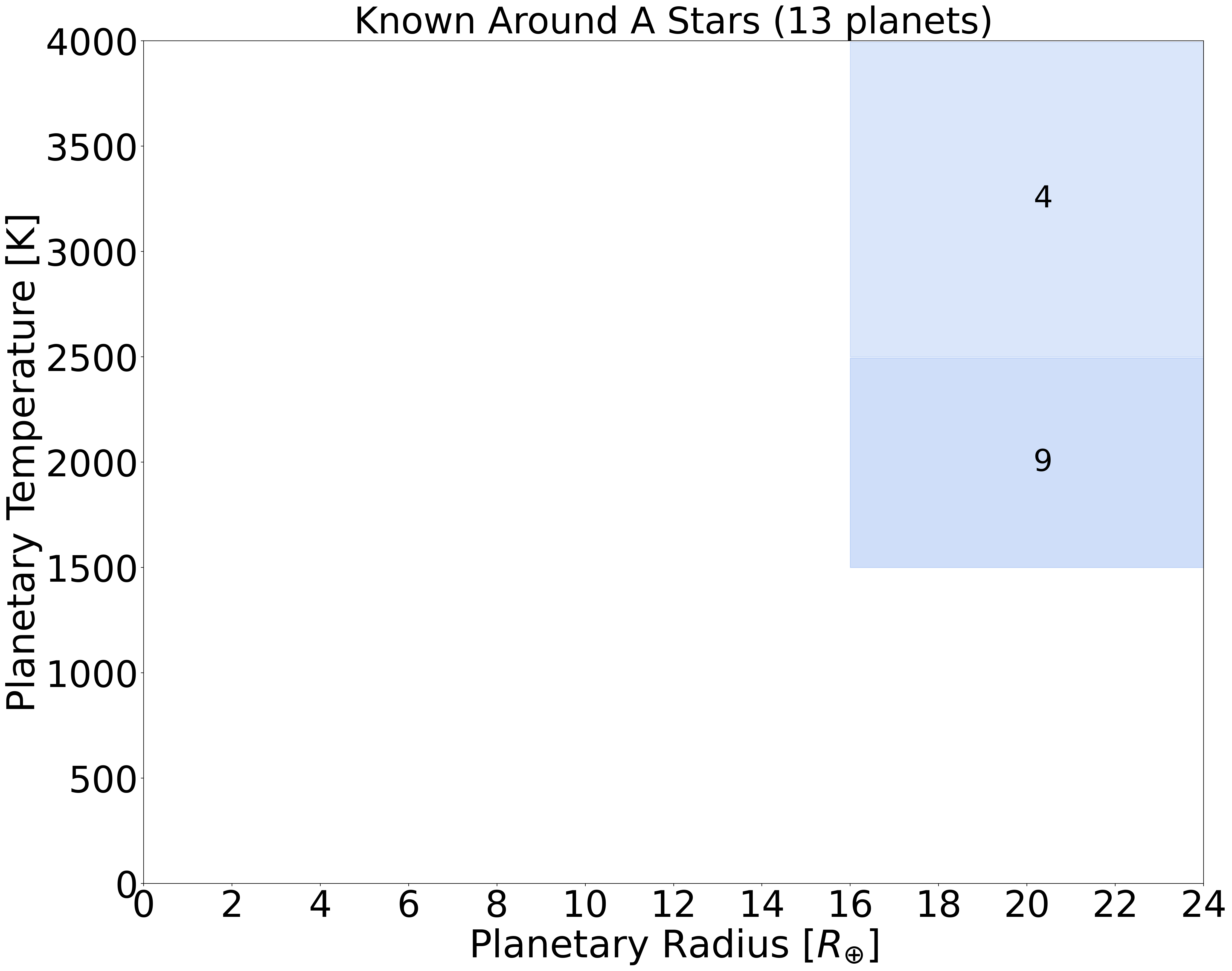}
    \includegraphics[width=0.45\textwidth]{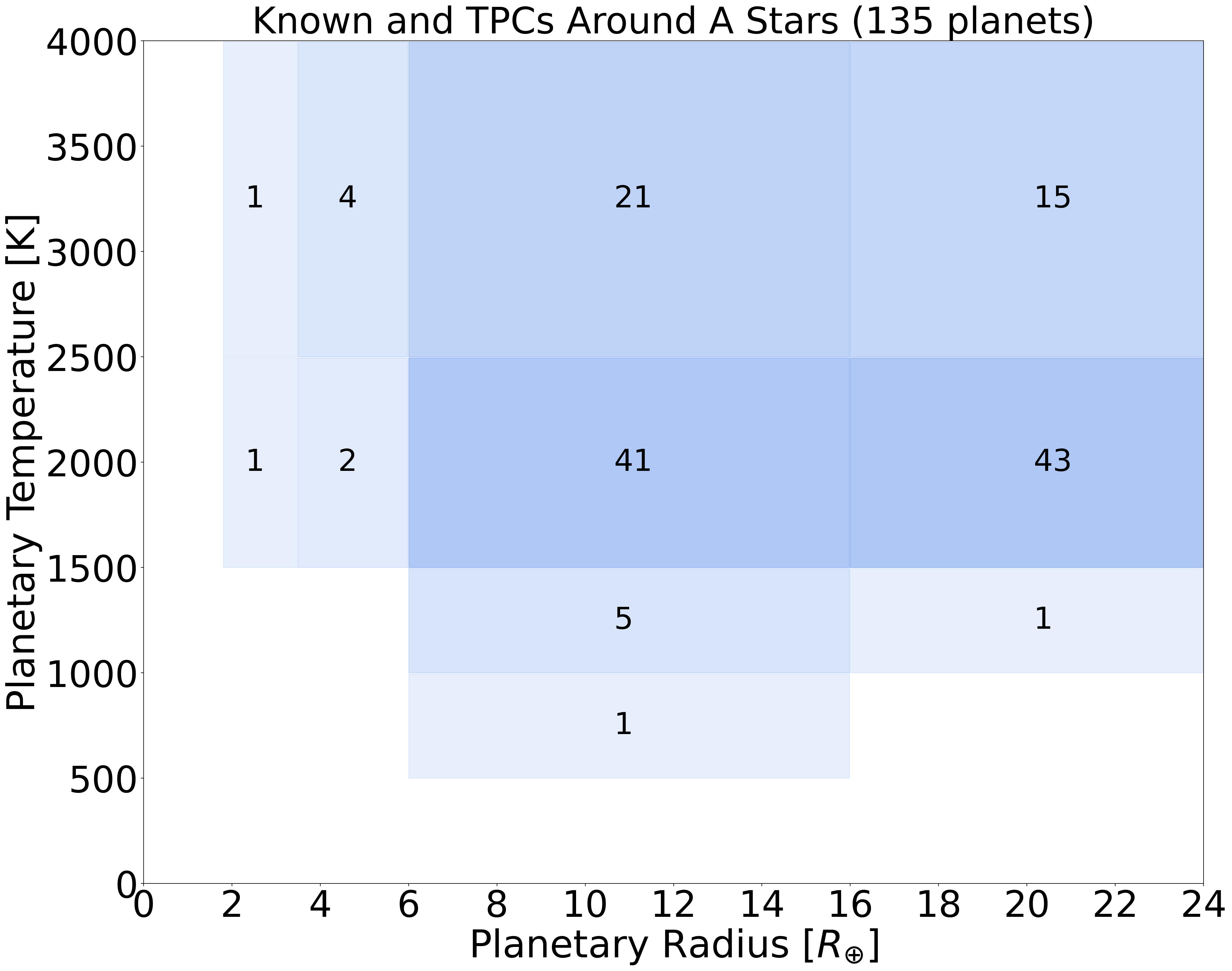}
    \caption{Temperature and radius distribution of known planets, including those found by TESS, (left) and these known planets in addition to the current TPCs (right), for F (top) and A (bottom) stars, that are suitable for Tier 1 study with Ariel.}
    \label{fig:distribution_fa}
\end{figure*}

\begin{figure*}
    \centering
    \includegraphics[width=0.45\textwidth]{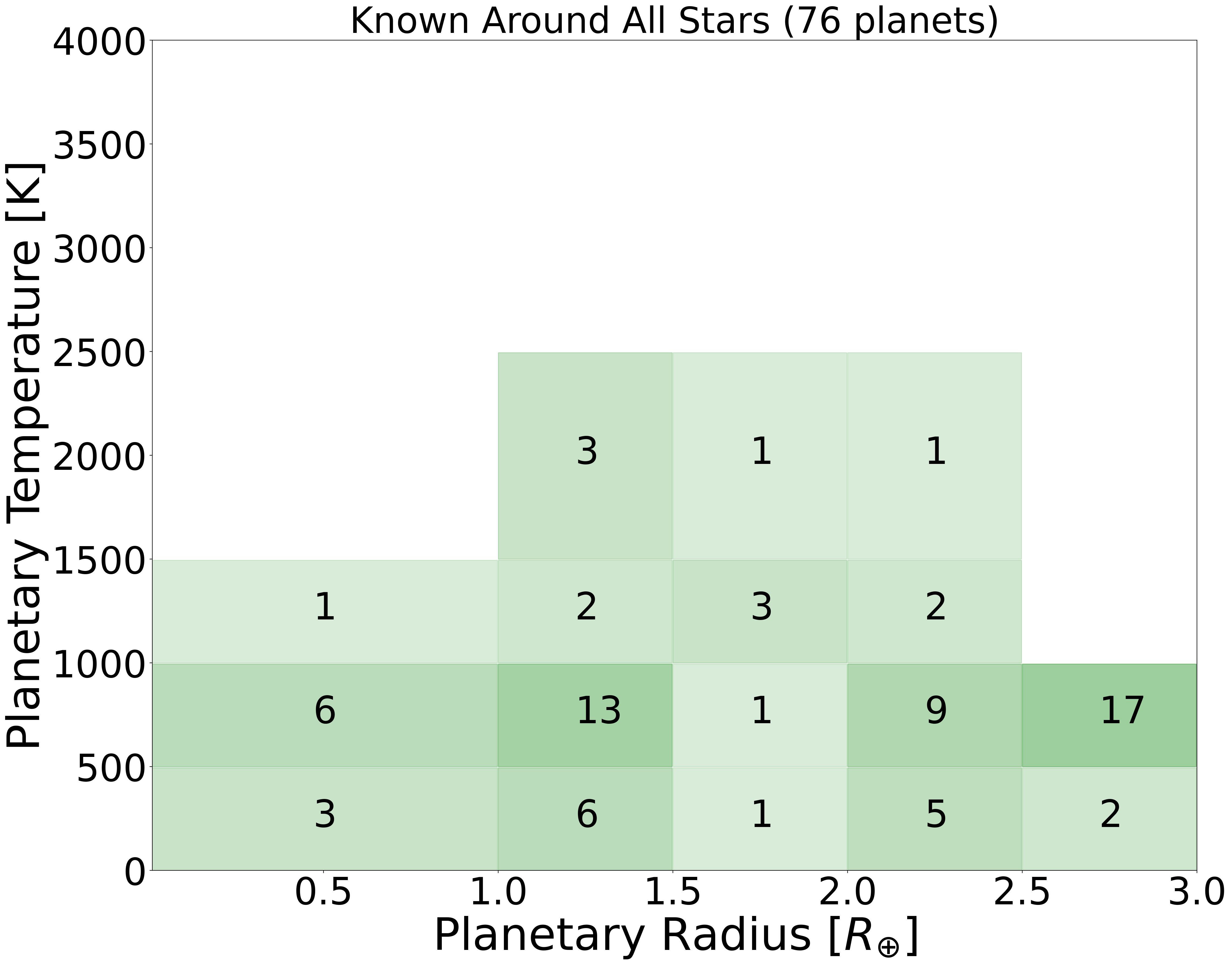}
    \includegraphics[width=0.45\textwidth]{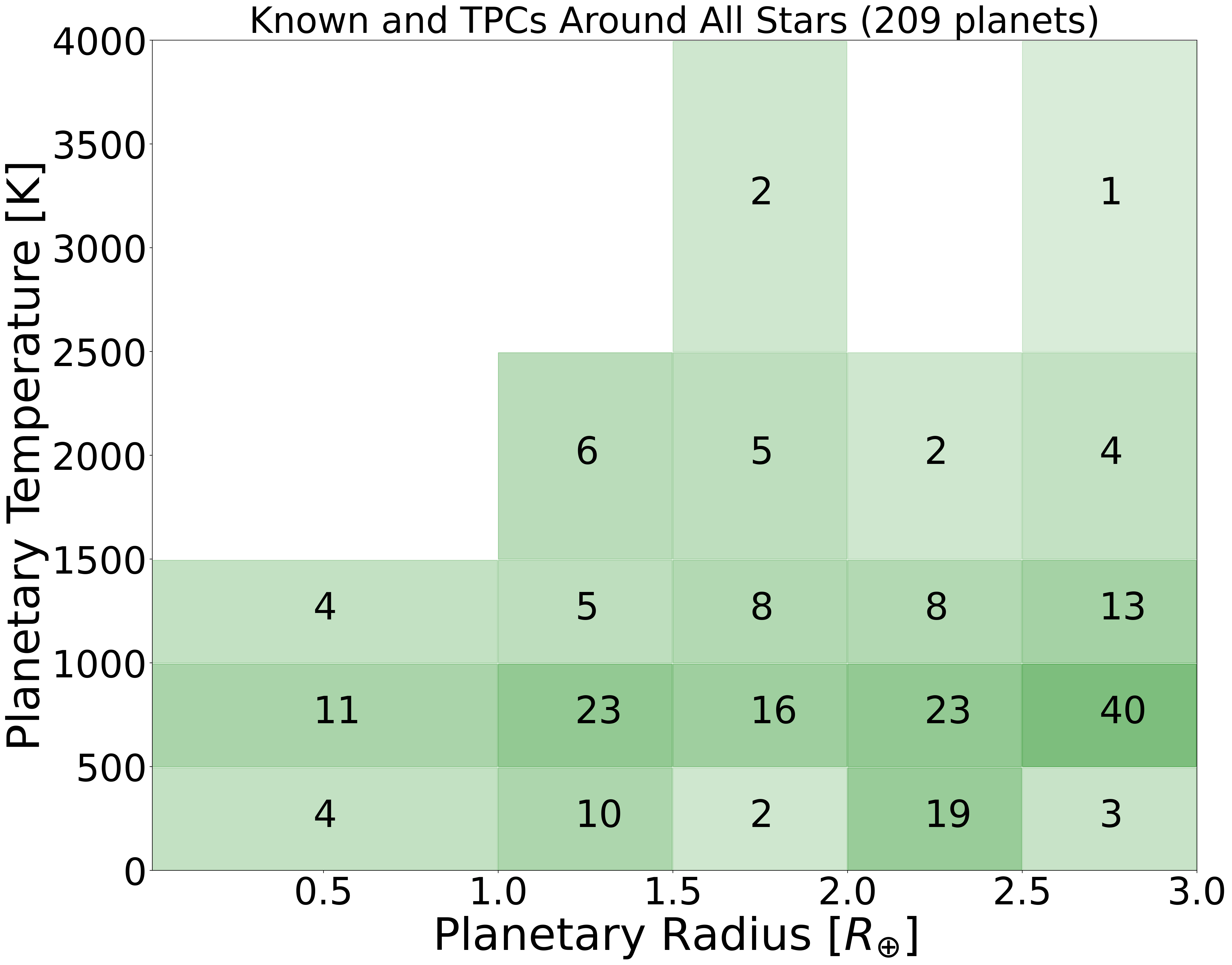}
    \caption{Temperature and radius distribution of known planets, including those found by TESS, (left) and these known planets in addition to the current TPCs (right), for worlds with radii smaller than 3 R$_\oplus$ that are suitable for Tier 1 study with Ariel.}
    \label{fig:distribution_small_pl}
\end{figure*}

\begin{figure*}
     \centering
     \includegraphics[width=0.95\textwidth]{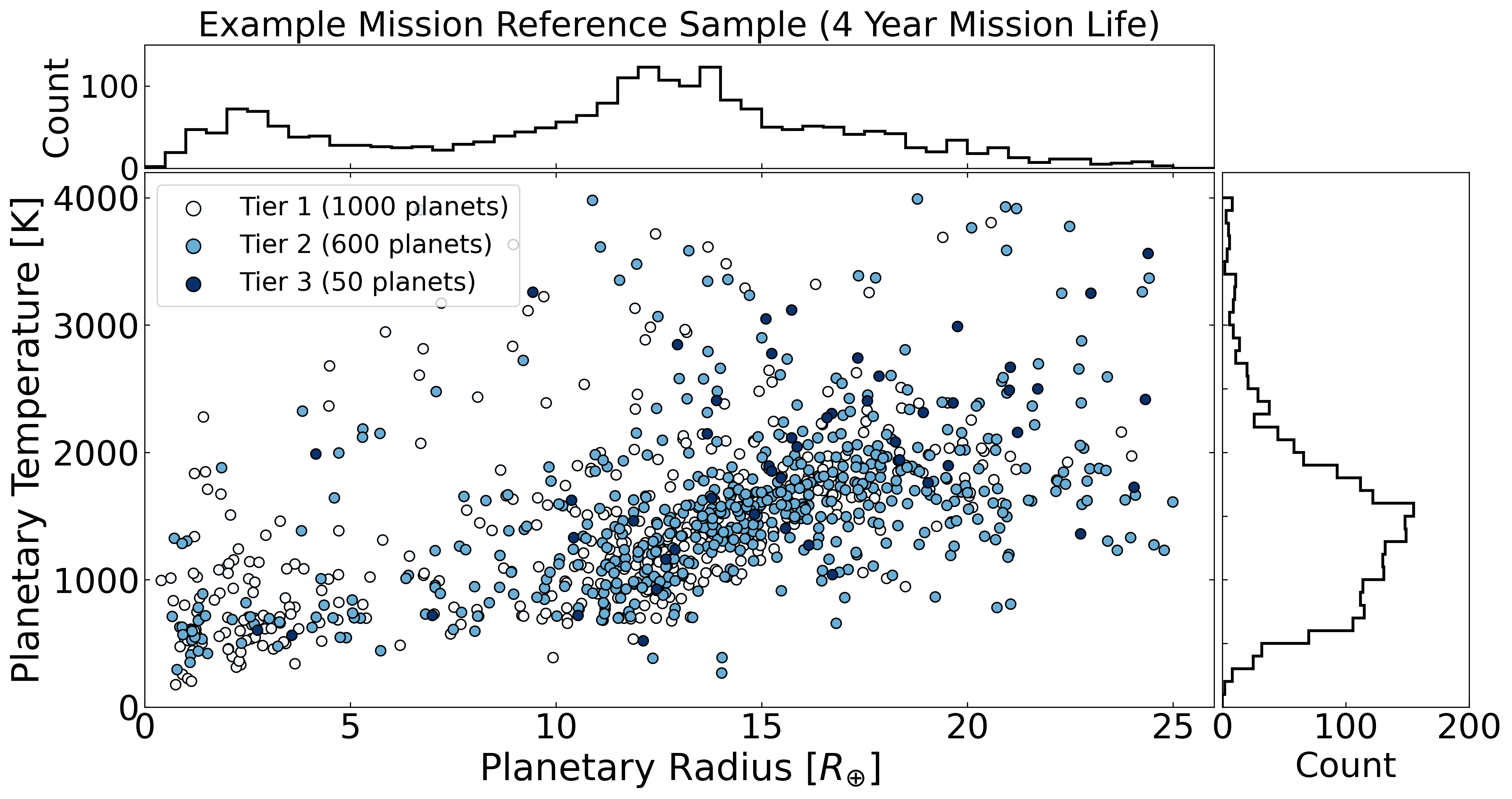}
     \caption{An example Mission Reference Sample (MRS) derived from currently-known planets and the TESS Planet Candidates (TPCs). The observing campaign shown would require 21,958 hours of telescope time, equivalent to 88.5\% of Ariel's available science time in the prime mission life ($\sim$24,800 hours). We note that around 10\% of time is expected to be left for Tier 4 observations and complementary science.}
     \label{fig:example_mrs}
 \end{figure*}
 
 \begin{figure*}
     \centering
     \includegraphics[width=0.95\textwidth]{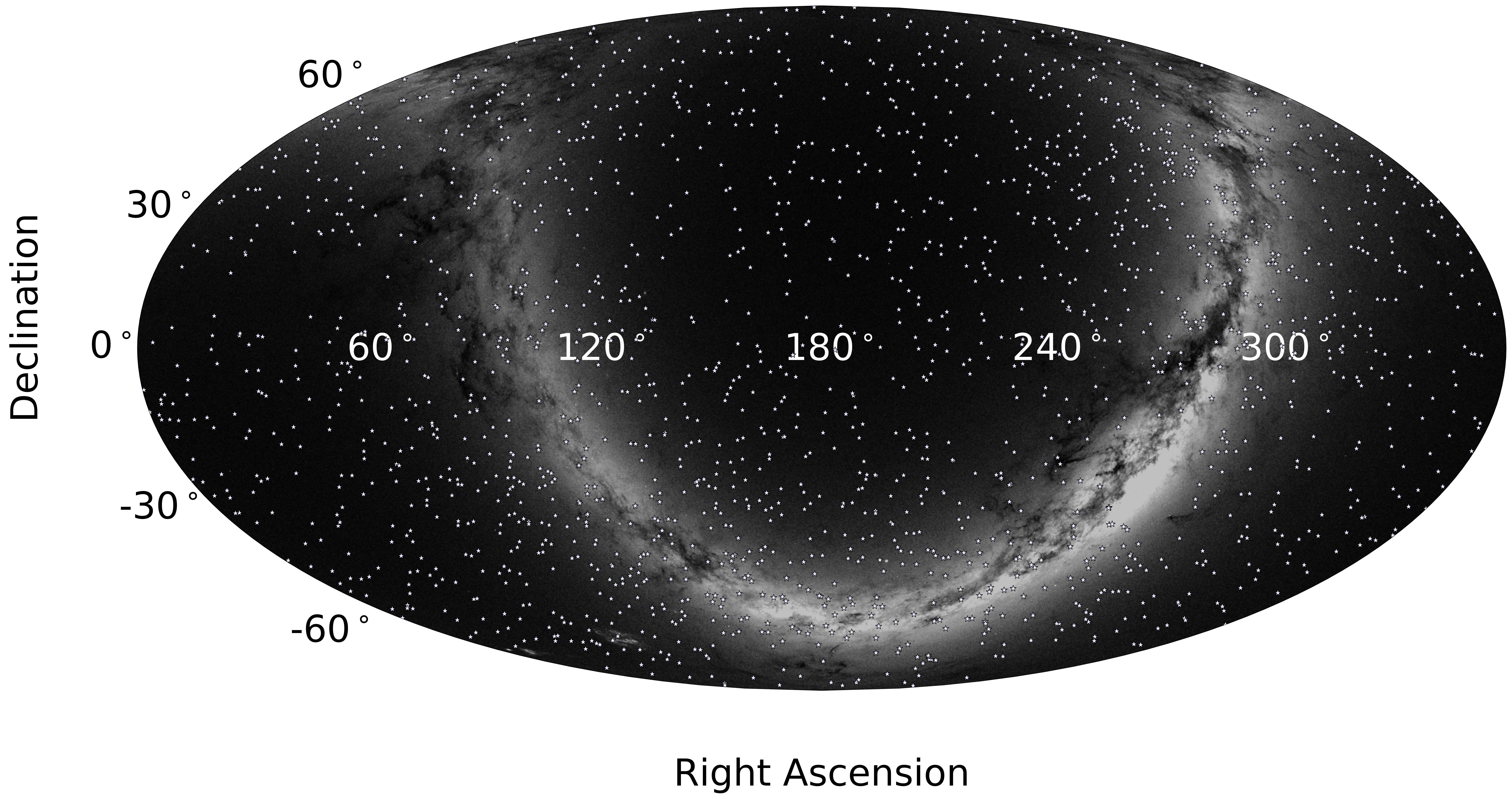}
     \caption{Sky locations of the example MRS, indicated by stars, with the brightest 35 million stars from the Gaia database \citep{gaia,gaia_dr2} plotted in the background. The host stars are spread across the sky, but with a lower density in the galactic plane. Transit searches here are harder due to the dilution of sources, but this will also be true of Ariel observations and thus targets in this plane may be less advantageous anyway.}
     \label{fig:example_mrs_sky}
 \end{figure*}

In Figure \ref{fig:distribution_mkg} and \ref{fig:distribution_fa}, the distribution of the potential Ariel targets are shown. In each plot, the left column shows the variety of planets available today, which includes those confirmed to date using data from TESS. In the right column, the known population has been combined with the TPCs. While we already known of a diverse sample of planets suitable for atmospheric characterisation, these plots highlight the impact TESS will have in increasing this variety yet further still. Furthermore, TESS will provide additional planets within each section of the parameter space to allow for more through comparisons between planets with similar bulk characteristics. We note that, in the right hand side of the Figures, false positives may still be present and the removal of these may reduce the overall diversity. For instance, signals indicative of planets with large radii on short periods have a relatively high false positive rate and, therefore, the potential expansion of the parameter space at the top right of each figure may not be real.

In \citetalias{edwards_ariel_tl}, we explored the ability of Ariel to study smaller planets. As these planets may have atmospheres which are not hydrogen-dominated \citep[e.g.][]{owen_wu,fulton}, or indeed no atmosphere at all, it is likely that a specialised observing plan will need to be devised to ensure they are studied in an efficient and thorough manner. While we do not go into depth on this here, in Figure \ref{fig:distribution_small_pl} we show the radius versus temperature distribution of small planets ($<3 R_\oplus$), highlighting the large number of potential targets as well as their diversity. Numerous parameters could affect the capability for a small planet to retain it's primordial envelope, including the stellar irradiation and planet mass \citep[e.g.][]{owen_wu,fulton,rogers_mass_loss}, and thus a wide range of targets will need to be studied if we are to understand which planets retain hydrogen-dominated envelopes, and which don't, as well as the pathways to secondary atmospheres. Hubble observations of Sub-Neptunes have had mixed levels of success in detecting atmospheres \citep[e.g.][]{kreidberg_gj1214,benneke_k2-18,tsiaras_k2-18,ares_iv} but spectroscopic observations of rocky planets have yet to yield convincing atmospheric detections \citep[e.g.][]{dewit_trappist,edwards_lhs,libby_gj1132,mugnai_gj1132}. The current lack of detailed atmospheric constraints for these worlds has increased the interest of the community in them. As such, a number of smaller planets will be studied as part of JWST's GTO and GO Cycle 1 programmes, with over half the planets observed being smaller than 2.5 R$_\oplus$. JWST will provide data with a wider spectral coverage and higher SNR than Hubble and, hopefully, these observations, and further proposals in future cycles, will provide the first detections of atmospheres around small, rocky worlds. Results from these studies will be crucial inputs into Ariel's strategy for observing smaller planets and the mission's potential for studying these worlds will be the subject of a focused manuscript in the future.

\subsection{An Example Mission Reference Sample}

As in \citetalias{edwards_ariel_tl}, we took the MCS, the list of potential Ariel targets, and created an example Mission Reference Sample (MRS), a selection of planets that could be observed in the prime mission life. In-keeping with previous works, we adopted the approach of aiming to choose a very diverse, and as complete as possible, combination of star/planet parameters while minimising the number of repeated observations by selecting the planets around the brightest stars. Again we classified planets using the bounds in Table \ref{tab:star_para} and ensure that, where possible, at least 2 planets within each star type/planet temperature/planet radius bin are contained within the MRS. We force 1000 planets to be observed in Tier 1 as well as 50 in Tier 3, each of which is assumed to be revisited 5 times in search of variability. We then fill the remainder of the mission time with Tier 2 observations, finding that, with the TOIs included in the sample, 600 planets could be observed in Tier 2 across the prime mission life under these conditions. We note that nominally Ariel will have 10\% of time its science time dedicated to Tier 4 targets and complementary science programmes. 

The distribution of the planetary radii and temperature of this example MRS is shown in Figure \ref{fig:example_mrs}, while the sky locations are given in Figure \ref{fig:example_mrs_sky}. Ariel target stars will be spread across the entire sky, hopefully helping to alleviate scheduling constraints \citep{morales_echo,morales_ariel}. The MRS derived here requires 21,944 hours, which is 88.5\% of Ariel's available science time in the prime mission life ($\sim$24,800 hours), leaving sufficient time for these other programmes. However, we also note that much of this Tier 4 time may be dedicated to phase curves. Naturally such observations would acquire transits and eclipses of the planets being studied and the planets which are suitable for phase curves studies with Ariel are also likely to be excellent targets for atmospheric spectroscopy in Tiers 1, 2 or 3. Therefore, this overlap means that primary science can be acquired from these Tier 4 observations, blurring the distinction between the two programmes and potentially opening up more time to conduct additional observations. In any case, the outcome of this study in clear: assuming a large portion of the planet candidates within the TOIs are true planets, will we already have a surplus of targets for Ariel and the mission will be capable of studying 1000 atmospheres during the prime mission life.

The Ariel mission should carry enough fuel for a mission extension to be possible. Hence we explored the impact on the number of planets which could be studied if the additional operating time were granted. We find that a two year extension would allow for 1400 planets to be studied in Tier 1 and 700 in Tier 2 across the entire mission life. Again we assumed roughly 10\% of time was dedicated to additional science observations. Such an extension would be an increase of 57\% in terms of science time and, from the currently derived target list, would yield increases of 40\% in the numbers of Tier 1 and 2 targets studied.


\begin{figure*}
    \centering
    \includegraphics[width=0.98\columnwidth]{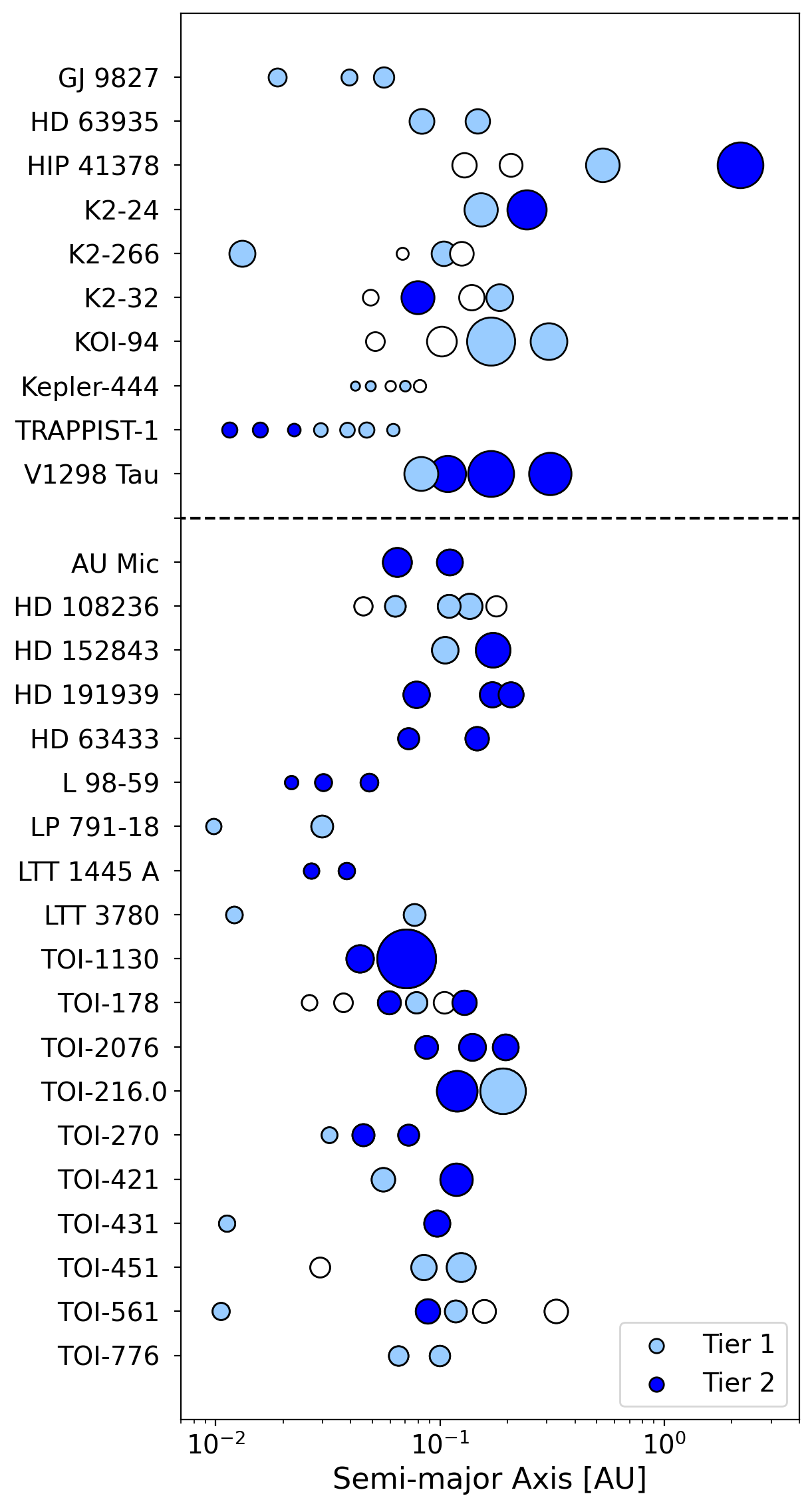}
    \includegraphics[width=0.95\columnwidth]{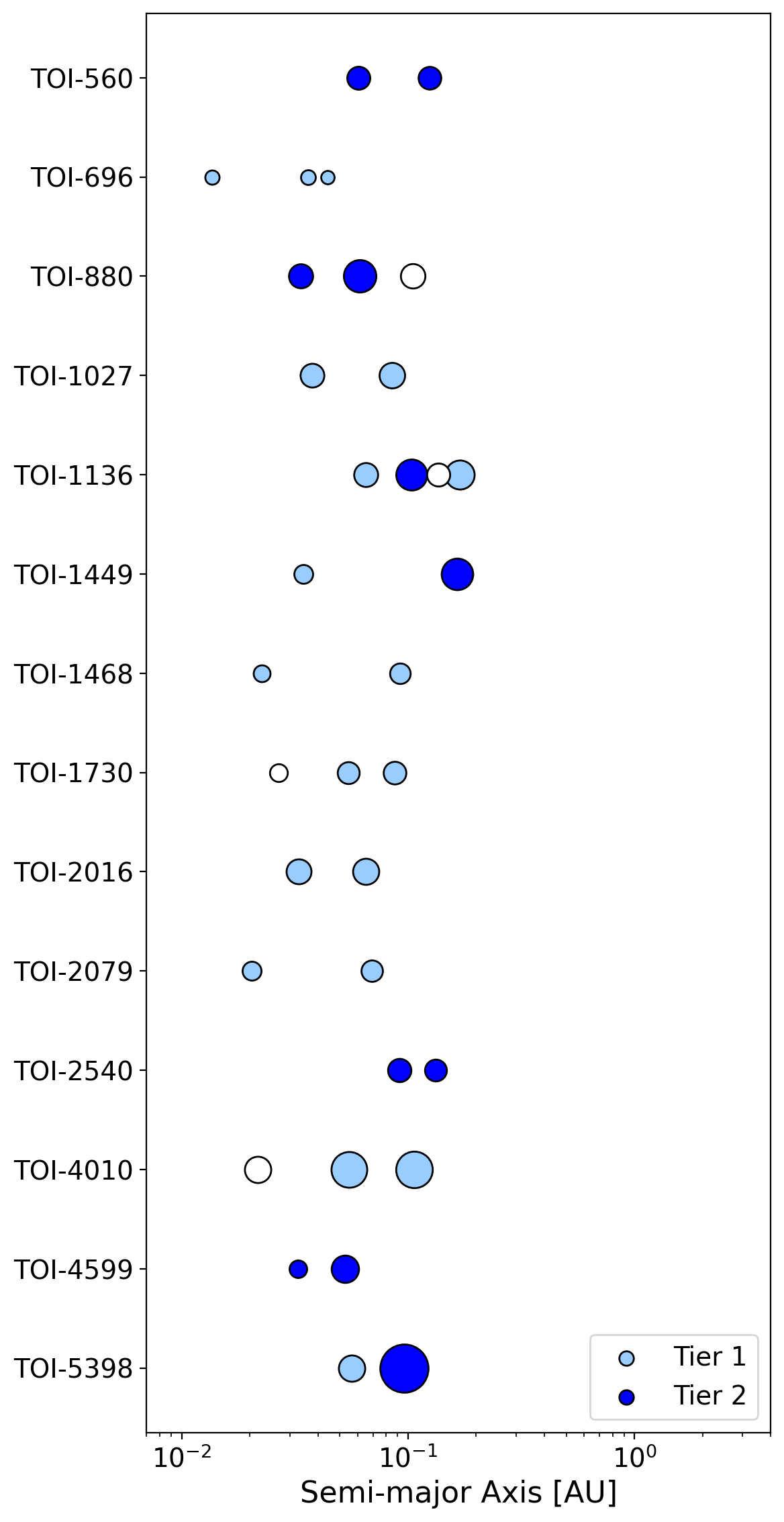}
    \caption{Known (left) and TPC (right) systems that host at least two worlds that could be studied in Tier 1 with Ariel. Colours indicate the suitability of characterisation (white: not suitable, light blue: Tier 1, dark blue: Tier 2) while the size of the circle reflects the planets radius. Those below the dotted line were listed as having been discovered by TESS in the NASA Exoplanet Archive.}
    \label{fig:multiple_planets}
\end{figure*}

\section{Systems which contain Multiple Planets for Atmospheric Study with Ariel}

While in general Ariel seeks to conduct comparative planetology across hundreds of targets, comparing the atmospheres of multiple planets within a single system could offer unique insights into their formation and evolution. Therefore, we isolated systems which had multiple planets that could be studied by Ariel, finding 31 known systems as shown in Figure \ref{fig:multiple_planets}. Within the TPCs, 17 such systems were found but we note that many multi-planet systems have already been confirmed with TESS \citep[e.g.][]{huang_toi1130,gunther_toi270,dawson_toi216,leleu_toi178} and thus are included within the known count. The large number of multiple planet systems that have been validated highlights their value as laboratories to study formation processes and, as TESS continues to provide additional data, one hopes further multi-planet systems will be found. 


We note there is disagreement between \citet{lacedelli_toi561} and \citet{weiss_toi561} about the number of planets in the TOI-561 system (4 and 3 claimed, respectively). As the NASA Archive lists all of these worlds, our methodology of creating a catalogue leads to 5 planets being listed for the system. For a study looking across the whole population, such an error should not affect the overall statistics. Planets b and c are consistently recovered by both studies and so the third planet found by \citet{weiss_toi561} is the only controversial world that is suitable for characterisation with Ariel within this system (the outer planets found by \citet{lacedelli_toi561} are deemed to take too much time to be studied based on the limits imposed here). Nevertheless, this example provides a warning that must be heeded as Ariel approaches launch and as the target list is further refined. The change in parameters between the TOI list and the confirmed catalogue is expected given the significant work that goes into confirming the planetary nature of the signal and characterising the system. However, it again provides an indication that one must be careful in the conclusions drawn from the TPC list and only once these systems are confirmed will we truly know there suitability for atmospheric studies with Ariel.

In light of this, we provide a first look at Ariel's capabilities to study multiple planets within one system by taking a confirmed system as an example: TOI-1130. The TOI-1130 system contains a warm Jupiter (R = 1.5R$_J$) on an 8.4 day orbit but, strangely, also contains an Neptune-sized world on a 4.1-day orbit \citep{huang_toi1130}. Systems with hot Jupiters rarely host other short period planets \citep{steffen_hj}, with any companions generally being at much larger orbital distances \citep{schlaufman_2016}. Given the brightness of TOI-1130 (K=8.351), these planets are ideal targets for atmospheric characterisation with Ariel and constraining their chemistry may provide indicators as to their formation history.

Models suggest that lower mass planets are incapable of accreting substantial gaseous envelopes, instead preferentially accreting higher-metallicity solids \citep[e.g.][]{mordasini_2012,fortney_2013}. As the primordial elemental abundances of giant, gaseous planets are expected to remain largely unchanged, measuring these can provide insights into the formation mechanisms of these planets. Studies of methane content of the gaseous planets within our own Solar System are in agreement with the predictions of the core-accretion scenario. By comparing the bulk characteristics of exoplanets to structural evolution models, there is evidence that a exoplanet mass-metallicity trend is likely but could differ for that seen in our own system \citep{thorngren_mass_met}.


Furthermore, elemental ratios are expected to a key indicator of where in the disc the planet formed. By constraining these ratios, such as that of carbon to oxygen, one may be able to uncover the formation and migration mechanisms governing the planet's evolution to its current state \citep[e.g.][]{oberg_co,turrini_ariel_2018,shibata_2020}. Shorter period planets, which will form the bulk of the population studied by Ariel, are likely to have formed far further out than they currently orbit, having migrated inwards over time \citep[e.g.][]{lin_migration,tanaka_mig}. The C/O ratio could be a tracer of whether the planet formed beyond the snow line; if the planet was originally outside the snow line it should have preferentially accreted carbon-rich gases leading to a high C/O ratio. Alternatively, if the planet accreted most of its material from inside the snow line, it should be more oxygen-rich and thus have a lower C/O ratio. However, modelling by \citet{turrini_formation} suggests that other elemental ratios, such as S/O or N/O, may provide a better opportunity to constrain where in the disc a planet formed. Nevertheless, as the C/O ratio is the most widely discussed ratio we stick to using it for this study.

We explored Ariel's capability to constrain these two key parameters, the C/O ratio and the metallicity, for the TOI-1130 system. We used the trend derived in \citet{thorngren_mass_met} to model the metallicities of TOI-1130 b and c. While \citet{thorngren_mass_met} found a trend between an exoplanet's mass and its metallicity, a stronger correlation was found when comparing the mass to the planet-to-star metallicity ratio. Hence we utilised their best fit model to this, which was
\begin{equation}
   \frac{Z_{P}}{Z_{S}}  = 9.7 M_{P}^{-0.45}
\end{equation}

where the metallicity of the star is determined from
\begin{equation}
    Z_{S} = 0.014 \times 10^{[Fe/H]}
\end{equation}

While TOI-1130 appears to be a metal-rich star, the Fe/H has not yet been measured, with current observations suggesting Fe/H $>$ 0.2 for the star \citet{huang_toi1130}. As in our case the value is a multiplicative term applied to both planets, we use Fe/H = 0.2. Furthermore, we note that the mass of TOI-1130\,b has not yet been constrained, with only a 3$\sigma$ upper limit of 0.17 M$_J$ placed upon it. Hence we utilised the relation from \citet{chen} to estimate the mass as 0.0407 M$_J$ (12.935 M$_\oplus$). We also note that the relation from \citet{thorngren_mass_met} did not extend to such small masses. Therefore, the trend within the TOI-1130 system, and within the population of Sub-Neptunes, may not reflect the one modelled here. However, we seek here only to present a proof of concept and leave a detailed study for future work.

\begin{table}[]
    \centering
    \begin{tabular}{ccc}\hline\hline
    & \multicolumn{2}{c}{TOI-1130} \\\hline
    T$_{\rm S}$ [K] & \multicolumn{2}{c}{4250$^\dagger$} \\
     R$_{\rm S}$ [R$_\odot$]  & \multicolumn{2}{c}{0.684$^\dagger$}  \\
     Fe/H  & \multicolumn{2}{c}{0.2$^*$}  \\ \hline
         & TOI-1130 b & TOI-1130 c  \\ \hline
     T$_{\rm P}$ [K] & 780 & 650 \\
     R$_{\rm P}$ [R$_J$]  & 0.326$^\dagger$ & 1.5$^\dagger$  \\
     M$_{\rm P}$ [M$_J$]  & 0.0407$^*$ & 0.974$^\dagger$  \\
     C/O  & 0.5 & 0.85 \\
     log(met) &  -0.041 & -0.662\\
     Cloud Pressure [Pa] & 1e3 & 1e3 \\
     No. Transits & 20 & 3 \\ \hline
    \end{tabular}
    \caption{Input parameters for the retrieval study of the TOI-1130 system. $^\dagger$denotes parameters taken from \citet{huang_toi1130}. $^*$denotes basic parameters that have not been yet measured and have been selected based on assumptions described in the text.}
    \label{tab:inputs_toi1130}
\end{table}

\begin{figure}
    \centering
    \includegraphics[width=0.9\columnwidth]{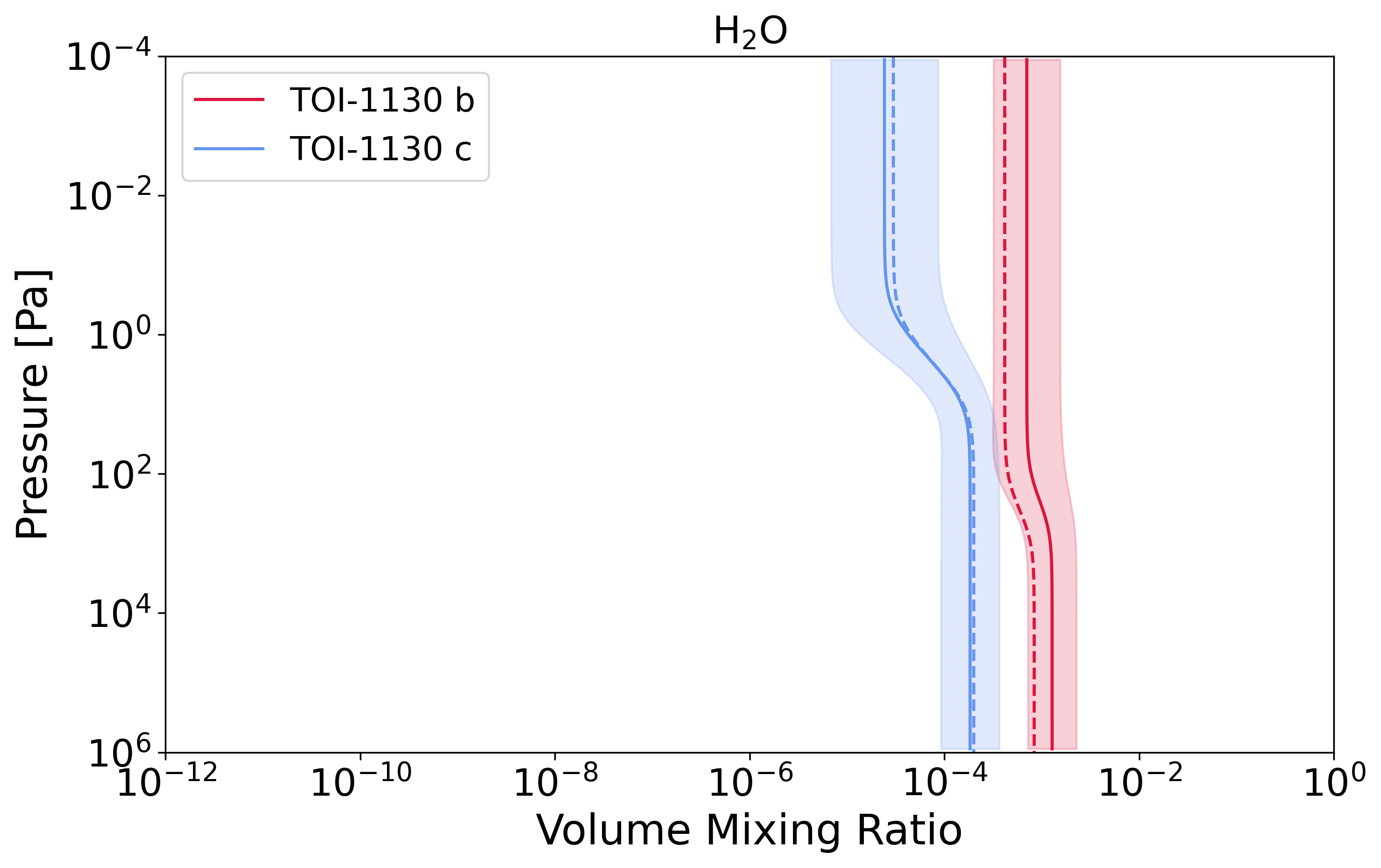}
    \includegraphics[width=0.9\columnwidth]{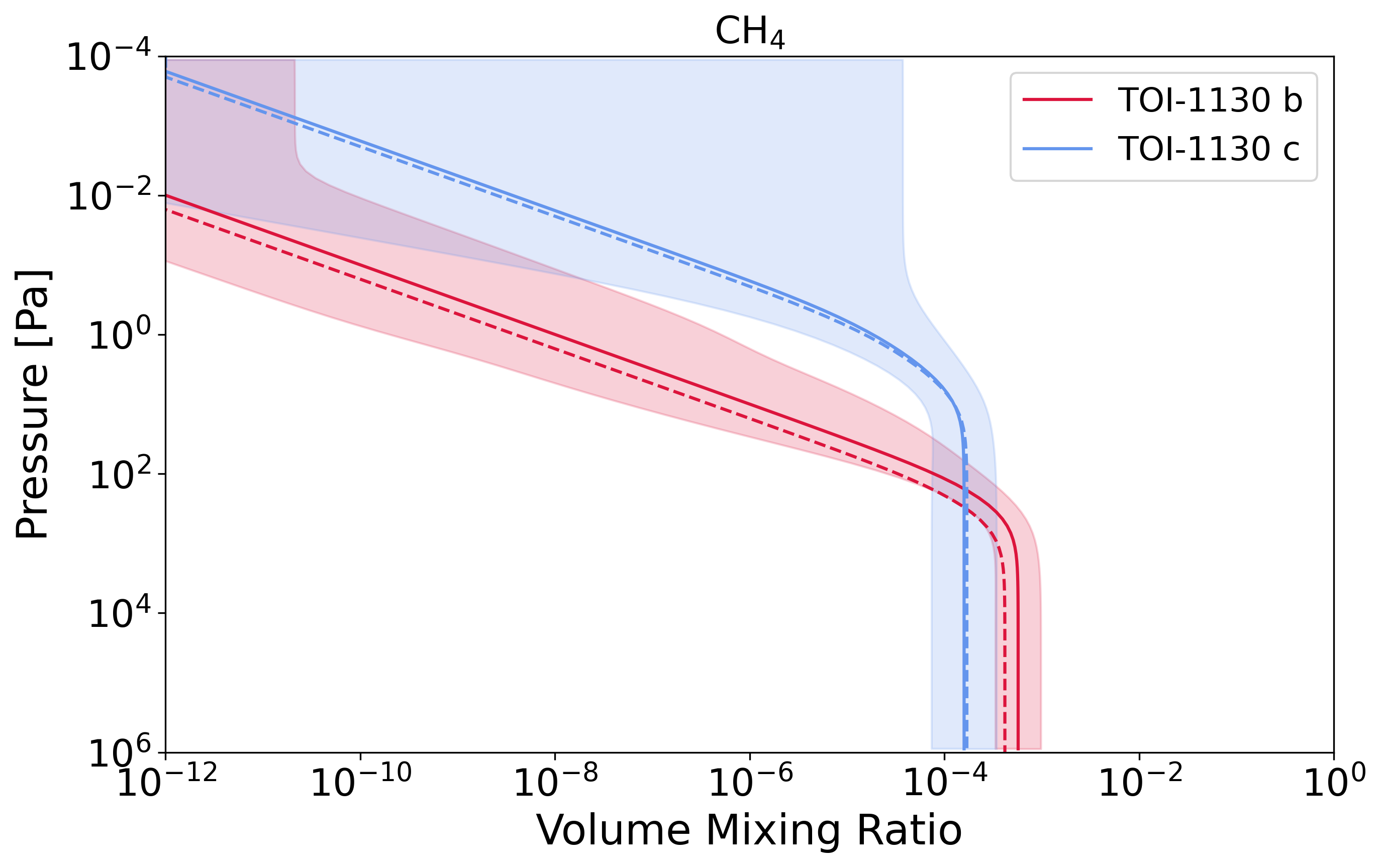}
    \includegraphics[width=0.9\columnwidth]{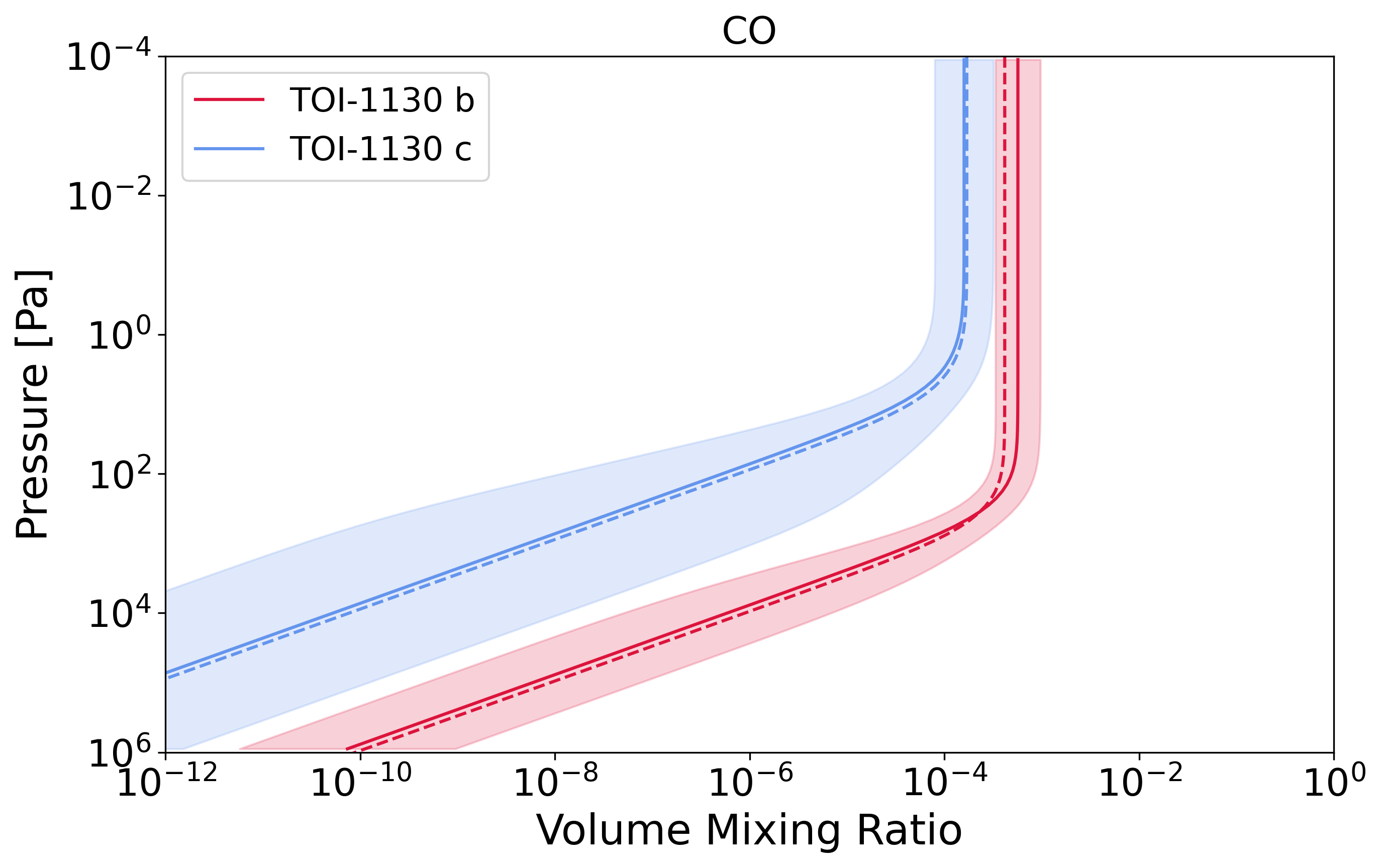}
    \includegraphics[width=0.9\columnwidth]{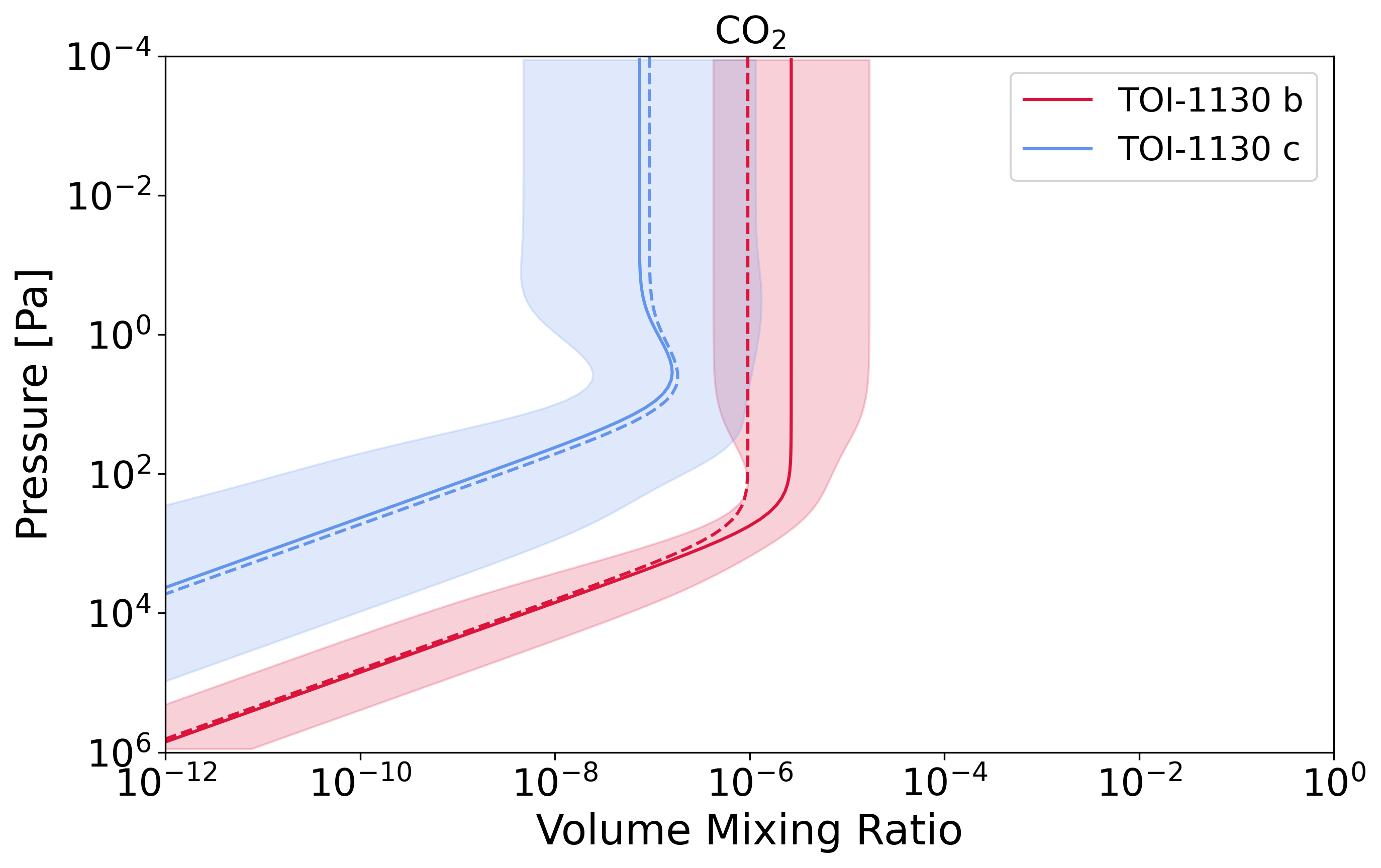}
    \caption{Constraints on the main carbon and oxygen bearing species from our retrievals for TOI-1130 b (red) and c (blue). The solid lines show the best-fit model while the 1$\sigma$ uncertainties are highlighted by the filled regions. The input values are given by the dotted lines. While the H$_2$O and CO abundances are well constrained, the CO$_2$ profile is less so, as is CH$_4$ in the upper atmosphere. Increasing the number of observations might improve our knowledge of these molecules and thus elemental ratios such as C/O. Ariel probes from the cloud-deck at 1e3 Pa up to roughly 1e1.}
    \label{fig:abundance_ret}
\end{figure}

\begin{figure*}
    \centering
    \includegraphics[width=0.945\textwidth]{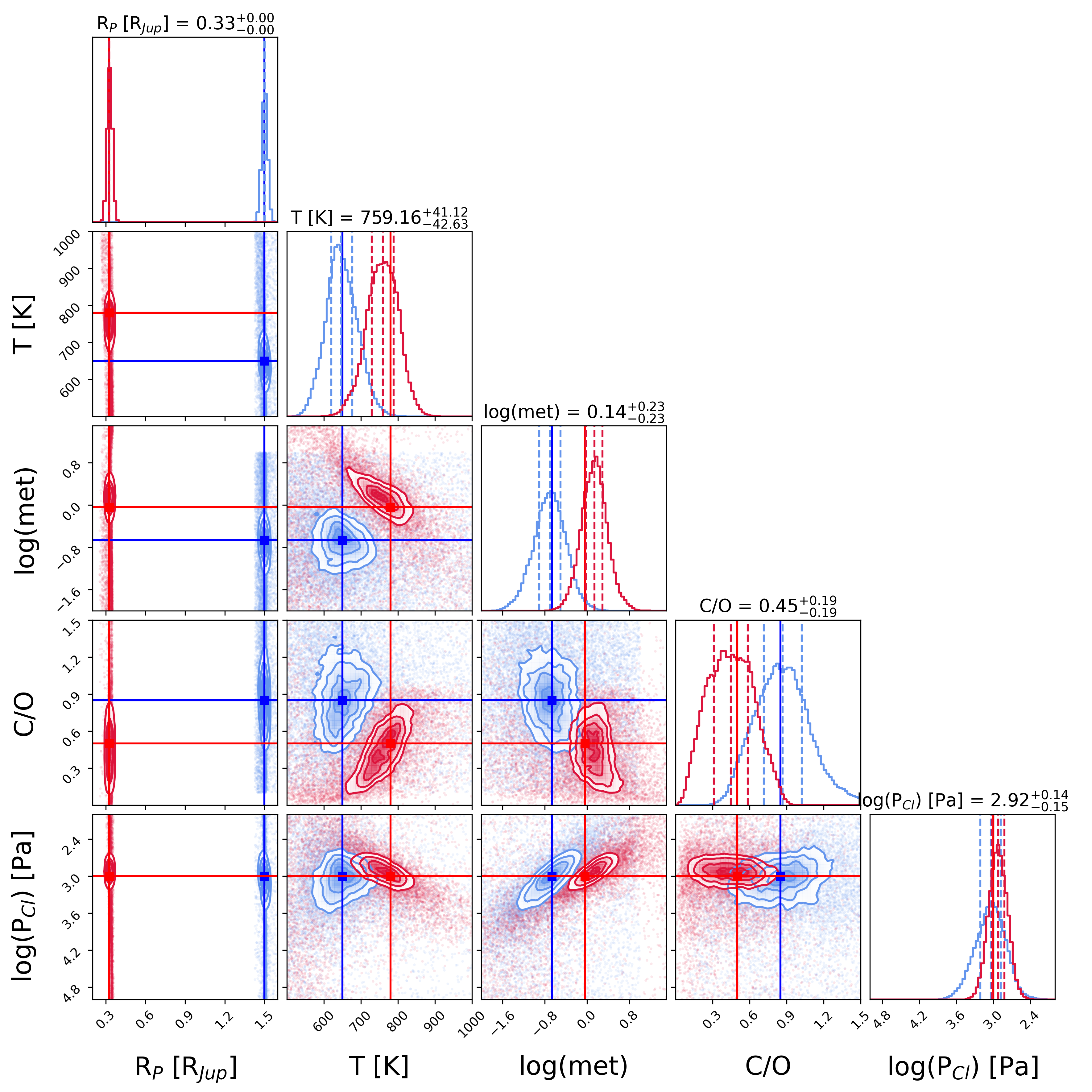}
    \caption{Posterior distributions for our retrievals of TOI-1130 b and c. For the assumed atmospheric metallicities and C/O ratios, Ariel would be able to distinguish between them and thus uncover a mass-metallicity trend in the system as well as inferring differences in their formation. }
    \label{fig:toi1130_co}
\end{figure*}

As well as modelling differing planet metallicities, we imposed different C/O ratios for the planets. We based these on the modelling in \citet{turrini_ariel_2018}, assuming a ratio of 0.5 and 0.85 for planets b and c respectively. We then modelled the spectra for these planets assuming equilibrium chemistry using the ACE package \citep{agundez_2012,venot_chem} which is a plugin for the TauREx 3.1 retrieval code \citep{al-refaie_taurex3,al-refaie_taurex3_chem}. We assumed isothermal temperature-pressure profiles and introduced a grey cloud deck at 5e2 Pa (0.005 Bar) and 1e3 Pa (0.01 Bar) for TOI-1130\,b and c, respectively, to produce spectral features in the HST WFC3 band which are equivalent in size to those seen in similar planets studied with this instrument \citep[e.g.][]{ares_iv}. We generated the error bars using ArielRad, assuming 20 observations for TOI-1130\,b and 3 for TOI-1130\,c, and our inputs are summarised in Table \ref{tab:inputs_toi1130}.

We find that, in the case presented here, the atmospheric constituents of TOI-1130\,b and c would generally be well constrained as shown in Figures \ref{fig:toi1130_co} and \ref{fig:abundance_ret}. We find that the metallicities of these planets could be distinguished (Figure \ref{fig:toi1130_co}), providing evidence for a mass-metallicity trend in the system if one should exist. Additionally, for the assumed C/O ratios, the solutions can be distinguished but less convincing than the metallicities. From Figure \ref{fig:abundance_ret} we note that CH$_4$ and CO$_2$ are less well constrained than H$_2$O and CO. Modelling a greater number of observations of TOI-1130 c in particular might help further distinguish the C/O ratios. However, we leave an in-depth analysis of this to a future detailed study of comparative planetology within multi-planet systems with Ariel, aiming here to only motivate the idea with a simple example and set forth the best systems with which to conduct such a study.


\section{Discussion, Conclusions and Future Work}

The Ariel target list, in the forms of an MCS and MRS, will continue to evolve as new systems are found, the instrument performance is refined, and the observing strategy is adjusted to maximise the science yield of the mission. We have shown here that TESS has already provided a plethora of planet candidates which are suitable for study with Ariel and, when combined with the currently-known population, its success will mean that the MCS could contain 2000 planets. 

\begin{figure}
    \centering
    \includegraphics[width=\columnwidth]{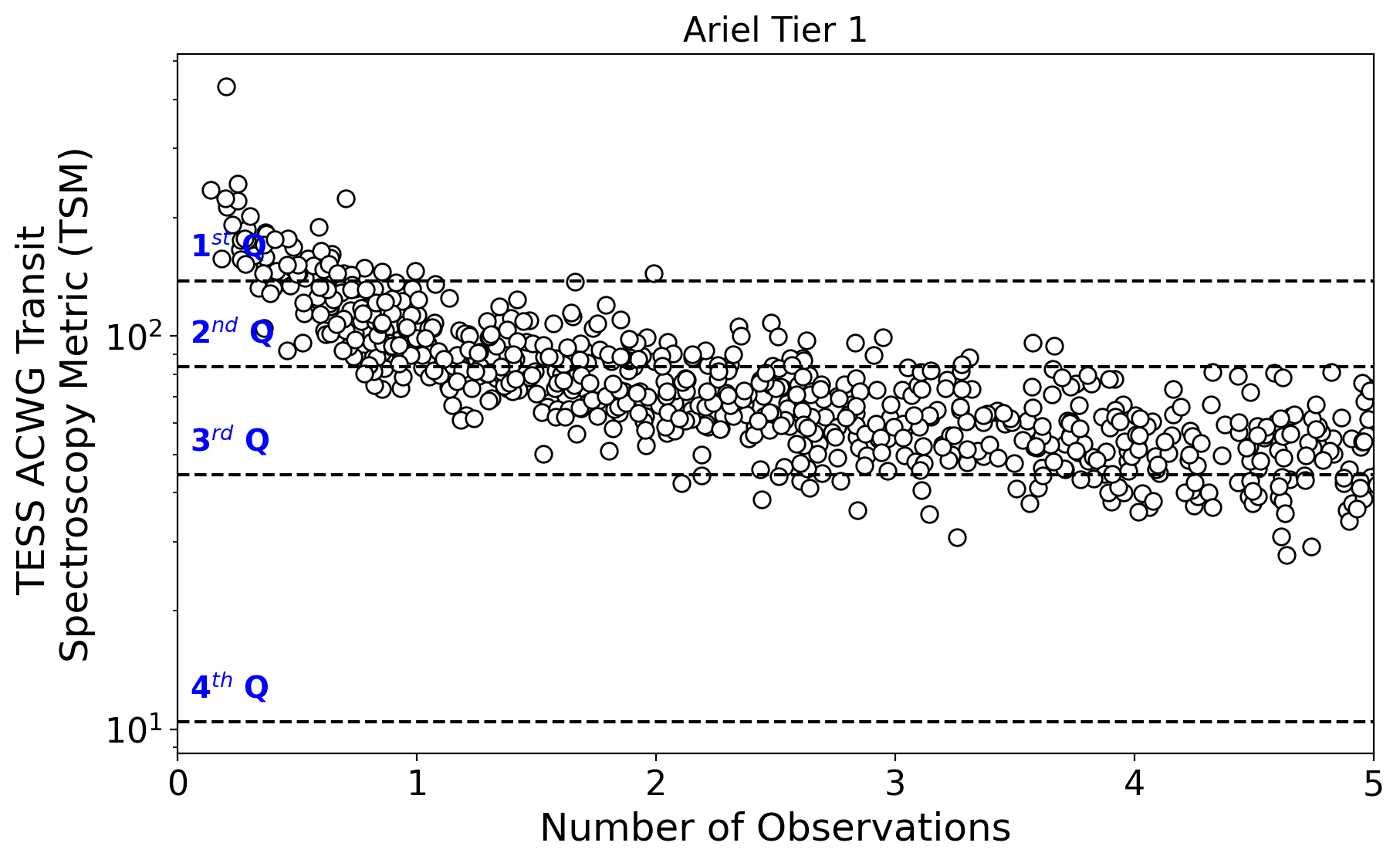}
    \includegraphics[width=\columnwidth]{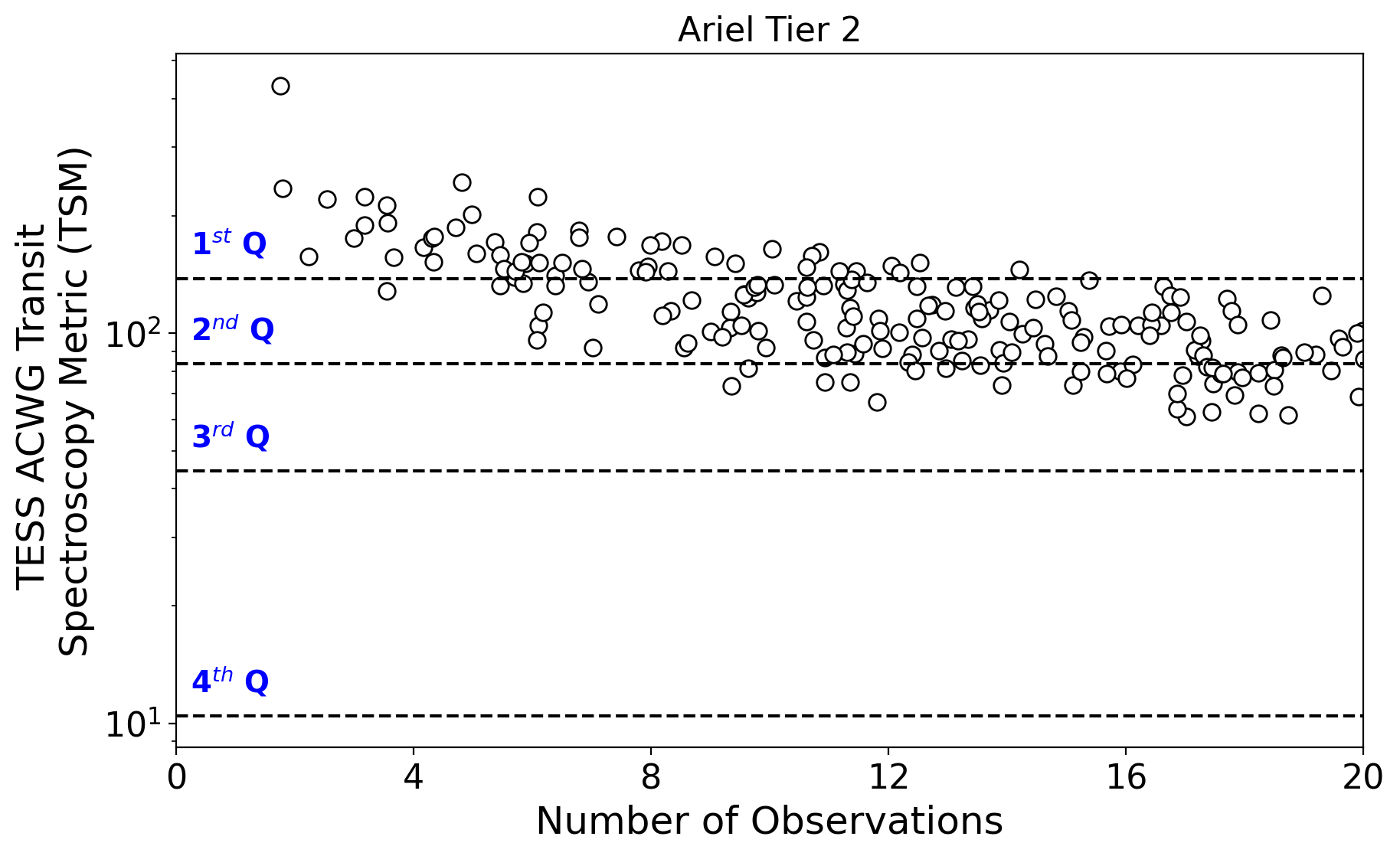}
    \caption{The transit spectroscopic metric, used by the TESS Atmospheric Characterisation Working Group (ACWG) to classify the suitability of atmospheres for characterisation \citep[TSM,][]{kempton_metric}, for TESS Planet Candidates (TPCs) that are potential targets for Ariel's Tier 1 and Tier 2 surveys. In Tier 1, Ariel will be capable of observing planets with far lower TSMs that the statistical sample considered in \citet{kempton_metric} (TSM$>$90), with some planets which have TSM$\sim$40 still being considered suitable for the mission. We note that the metric was designed only for JWST NIRISS and, therefore, is not a robust method for assessing the suitability for study with Ariel.}
    \label{fig:metric}
\end{figure}

While the majority of these candidates still require confirmation, the continued persistence and devotion of those involved in following up these potential planetary signals, as well as the further extension to the TESS mission, will ensure a large and diverse population of planets are available for Ariel to characterise. The metrics derived in \citet{kempton_metric} are being widely used to direct the TESS follow-up of potential atmospheric targets. In Figure \ref{fig:metric}, one can see the number of Ariel observations needed to reach the Tier 1 and Tier 2 requirements plotted against the Transit Spectroscopy Metric (TSM) from \citet{kempton_metric} for the TPCs analysed here. Over-plotted are the suggested quartiles for larger planets from \citet{kempton_metric}, with the scaling factor removed. \citet{kempton_metric} suggested that the cut-off TSM for a statistical sample of gaseous planets should be placed at around 90, a value often quoted and benchmarked against in TESS discovery papers. While this value seems applicable to Ariel Tier 2, we suggest that the boundary should be lower for Tier 1, with Ariel being capable of potentially studying planets with a TSM$\sim$40 or above. However, we note that this metric was designed specifically for JWST NIRISS and thus utilises the star's magnitude in the J band. If one takes two stars which have the same J band magnitude but different temperatures, the stellar flux at longer wavelengths will be greater for the cooler star. Therefore, as Ariel's science requirements are driven by the performance at longer wavelengths (1.95-7.8 $\mu$m), its use in this context may not be valid and the performance for planets around cooler stars may be under-represented with respect to those around hotter ones. Hence, we recommend the use of ArielRad \citep{mugnai_ar} to truly assess the suitability of a target for study.

In addition to the multitude of TESS discoveries, the sustained analysis of Kepler/K2 data \citep[e.g.][]{castro_k2,leon_k2,valizadegan_kep_ml,zink_k2}, ongoing ground-based surveys \citep[e.g.][]{wheatley_ngts,sebastian_spec}, and ESA's PLATO mission \citep{rauer} will also add to the population of exoplanets from which the Ariel sample will be derived. Moreover, other space-based facilities, such as CHEOPS \citep{benz_cheops} and Twinkle \citep{edwards_exo}, can be expected to find a handful of targets by searching for transits of planets detected by radial velocity \citep[e.g.][]{delrez_cheops} or conducting observations of systems with transiting worlds in the search of additional bodies \citep[e.g.][]{bonfanti_cheops}.


Such an overabundance of potential targets is undoubtedly beneficial, yet it offers many challenges. While much of the final MRS will be selected based upon which targets occupy the most sparsely-populated regions of parameter space or are best placed to answer key questions on the nature of exoplanet atmospheres, prioritisation may also be based upon how well-characterised the planets bulk parameters and host star are. If we are to seek trends between atmospheric composition and the bulk characteristics of planets, we must first know both of these. While Ariel will give us the former, the latter knowledge must generally be acquired before launch to ensure the sample of planets selected allows for these comparisons to be drawn. The key parameters needed include the stellar metallicity and the planet's mass and transit ephemerides.

Luckily, a large cohort of researchers\footnote{The Ariel consortium is currently composed of over 500 scientists.} are working to ensure this is the case. For instance, stellar parameters such as age and metallicity must be known and work is underway to homogeneously derive such parameters \citep{brucalassi_ariel,danielski_ariel,magrini_ariel_star}. Furthermore, the ephemerides of the potential planets must be well known to ensure efficient scheduling. Here again, work is underway to refine the periods of these worlds using citizen science, with thousands of light curves being observed by amateur facilities as part of the ExoClock project \citep{kokori_exoclock,kokori_exoclock2} and by secondary school students through the ORBYTS programme \citep[e.g.][]{edwards_orbyts_iii}. Planets with non-linear ephemerides, such as those within multi-planet systems that experience transit timing variations (TTVs) due to the gravitational interaction of the planets, will require particular care and attention \citep[e.g.][]{dawson_toi216,kipping_toi_216,ducrot_trappist}. By highlighting these key systems in Figure \ref{fig:multiple_planets}, and continuing to update this list, we hope to motivate such follow-up in a timely manner.

Accurately knowing the planet's mass is also useful, particularly for smaller or cloudier worlds \citep{changeat_mass,batalha_mass}. Several studies have explored the time need to provide radial velocity measurements necessary to constrain planet masses for Ariel \citep[e.g.][]{barnes_mass,demangeon_rv} and the ongoing work of radial velocity teams, both within the Ariel consortium and outside of it \citep[e.g.][]{lillo_box_k2,chontos_keck,nielsen_toi125,kaye_toi270,van_eylen_toi270}, are ensuring that the best targets for atmospheric characterisation are followed-up. Nevertheless, it may not be possible to conduct detailed follow-up of all the TPCs which may be of interest to the Ariel mission. 

The formation of the Ariel MRS will also need to account for the knowledge gained through previous spectroscopic studies of exoplanetary atmospheres. Ground-based and space-based spectroscopy are constantly providing new insights into exoplanetary atmospheres. High-resolution spectroscopy from the ground is becoming ever-more fruitful \citep[e.g.][]{pino_k9,tabernero_w76,giacobbe_hr,wardenier_w76}. At lower resolutions, transit and eclipse studies with ground-based facilities and Hubble continue to deliver new views of atmospheres \citep[e.g.][]{mcgruder_w31,braam_w31,saba_w17,yip_w96}, with a recent focus on smaller planets \citep[e.g.][]{dewit_trappist,benneke_k2-18,tsiaras_k2-18,diamond_lowe_lhs1140,edwards_lhs,gressier_t1h}, and findings from these studies will further our understanding of exoplanetary atmospheres. 

Through its GTO and Cycle 1 GO programmes, a significant amount of time has been attributed for JWST to study exoplanets via transit and eclipse spectroscopy. The high precision data that will be acquired, on top of a wider wavelength coverage with respect to current space-based instruments, will unlocking new avenues for exoplanet characterisation \citep[e.g.][]{changeat_k218,phillips_jwst,pidhorodetska_jwst}. The lessons learned from these studies, particularly those of smaller worlds, will influence the choices made to ensure the scientific yield of Ariel is maximised.

\begin{figure}
    \centering
    \includegraphics[width=\columnwidth]{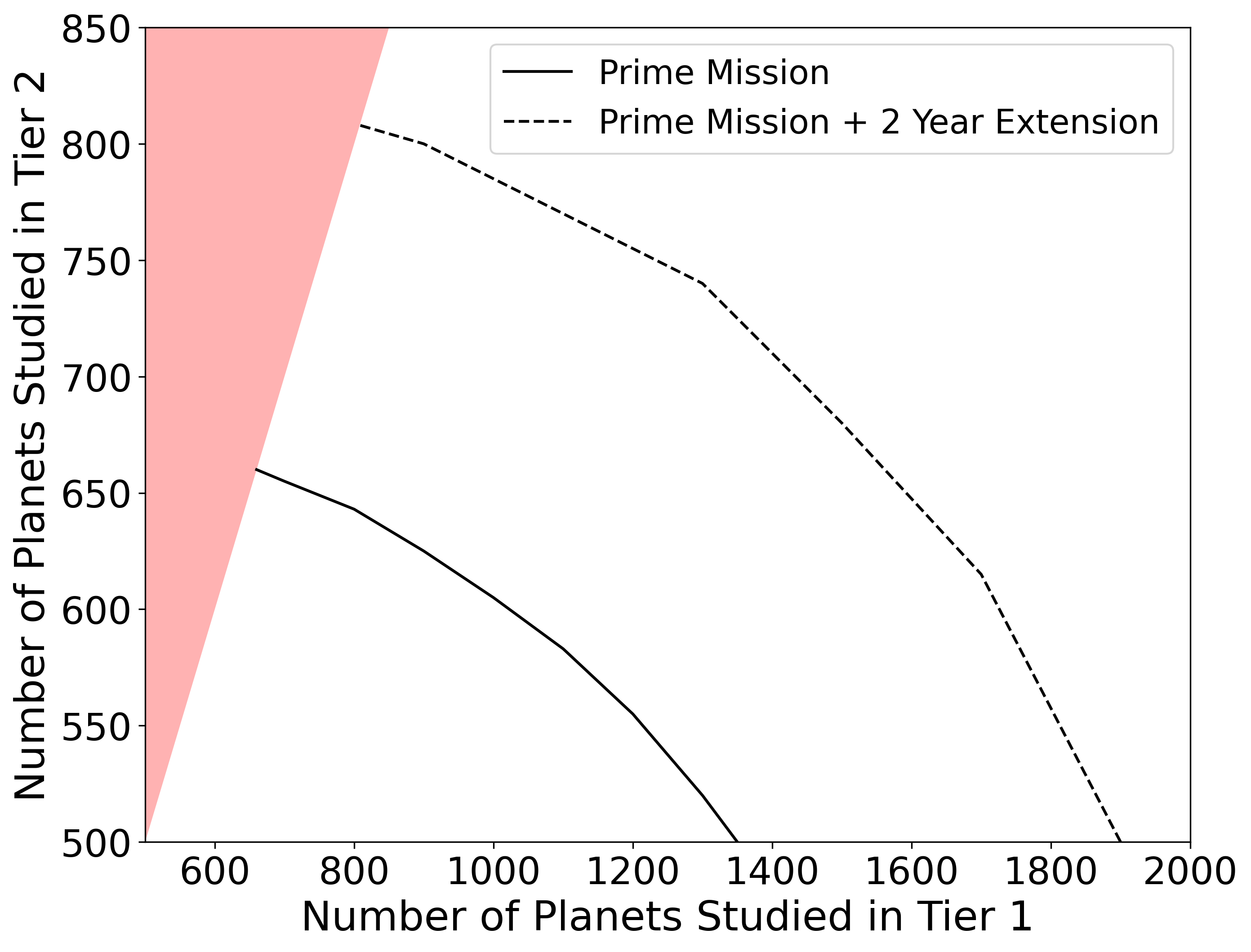}
    \caption{The relationship between the number of Tier 1 and Tier 2 planets that can be observed during the prime mission life. The red region is forbidden: by their very definition one cannot observe more targets in Tier 2 than in Tier 1. We also show what could be achieved with a 2 year mission extension. In all cases, 50 planets are assumed to be studied in Tier 3 and 10\% of Ariel's science time is left for Tier 4 observations and complementary science programmes.}
    \label{fig:t1_t2_trade}
\end{figure}

While thus far we assumed 1000 planets would be studied in Tier 1, this is by no means fixed. We varied the number of Tier 1 planets studied, experimenting with the impact this would have on the number of targets that could be observed in Tier 2. Assuming that at least 500 planets are required in Tier 2, it could be possible to study around 1350 planets from the list derived here during the primary mission life. Reducing the required number of Tier 1 targets increases the time available for Tier 2 but the returns are quickly diminishing. The optimal strategy, purely from an efficiency perspective, occurs when the change in the gradient is sharpest. From Figure \ref{fig:t1_t2_trade}, this is currently at around 1050 Tier 1 planets, which corresponds to 595 Tier 2 targets if one assumes 10\% of Ariel science time is reserved for Tier 4 and complementary science. These numbers are based on a four-year primary mission life but if a two year extension were granted, and the time utilised to extend the observing strategy of the primary mission life, we could potential study a far greater number of worlds. However, the results from the primary mission may suggest alternative lines of inquiry. The tiering system of Ariel has been designed to account for our general lack of knowledge around the properties of exoplanet atmospheres and their correlation with the bulk characteristics of a system. Given the primary mission will provide us with the first demographical study of exoplanet atmospheres, the extended mission could then be utilised to study specific populations of interest. The strategy could involve delving deep into trends uncovered in the prime mission, giving extra time to targets or correlations which weren't fully explored, conducting more time intensive observations or pursuing observing strategies that are high risk/high reward.

In reality, the Ariel Mission Consortium (AMC) and the Ariel Science Team (AST) will decide the observing plan of Ariel based upon the scientific yield which is dependent upon a number of parameters, many of which are qualitative in nature. Studies of the impact of altering the number of planets in each Tier will be required, with population-level studies needed to ascertain the science loss/gain from adopting different strategies \citep[e.g.][]{changeat_alfnoor,mugnai_alfnoor}. While not discussed here, Ariel has a fourth Tier to account for targets or observations which do not fit into the main structure of the survey. Much of this time might be dedicated to exoplanet phase curves, observing a planet throughout its entire orbit \citep[e.g.][]{stevenson_w43,changeat_phase,dang_xo3,may_w76}, and studies are underway to find the best targets for phase curve studies with Ariel \citep{charnay_phase,moses_phase}. Additionally, this time could be utilised to observe small, rocky worlds which may host secondary atmospheres, as outlined in \citetalias{edwards_ariel_tl}. Further in-depth studies of Ariel's capabilities across specific populations will be needed \citep[e.g.][]{encrenaz_ariel_temp,ito_rocky_pl} to provide meaningful constraints on the potential science yield. These studies, alongside community engagement, will act as guides for the AMC and AST as they construct an ideal MCS. The final MCS will be endorsed by the AST and reviewed under the responsibility of the ESA Advisory Structure before launch. More details on this procedure can be found in the Ariel Science Management Plan.

The confirmed planets and TESS planet candidates that are considered potential targets for Ariel using the methodology described here are given in Tables 3 and 4. To guide the studies described above, and facilitate engagement with the wider community, an Ariel Target List website, which will be maintained by the AMC, is being constructed (Al-Refaie, in prep). The site will allow users to keep track of all the planets that are potential viable targets for Ariel, motivating further preliminary follow-up as well as theoretical studies into Ariel's capabilities. Through this collaborative platform, we hope to ensure the Ariel mission delivers datasets of value to the entire extrasolar community and that their interests are reflected in the final observing strategy.

\section*{Acknowledgements}

BE is a Laureate of the Paris Region fellowship programme which is supported by the Ile-de-France Region and has received funding under the Horizon 2020 innovation framework programme and the Marie Sklodowska-Curie grant agreement no. 945298. This project has also received funding from the European Research Council (ERC) under the European Union's Horizon 2020 research and innovation programme (grant agreement No 758892, ExoAI) and from the Science and Technology Funding Council (STFC) grant ST/S002634/1.

We acknowledge the availability and support from the High Performance Computing (HPC) platform DIRAC, which provided the computing resources necessary to perform this work. The research conducted here has made use of the NASA Exoplanet Archive, which is operated by the California Institute of Technology, under contract with the National Aeronautics and Space Administration under the Exoplanet Exploration Program. Additionally, this paper relied upon data products from by the TESS mission, funding for which is provided by the NASA's Science Mission Directorate. We acknowledge the use of public TOI Release data from pipelines at the TESS Science Office and at the TESS Science Processing Operations Center.

We thank the Ariel Science Team, and many members of the Ariel consortium, for their feedback on this manuscript throughout its development. We are also indebted to Lorenzo V. Mugnai for his ongoing efforts to develop and maintain the ArielRad tool, which facilitated this work. Finally, we thank the referee for their insightful comments and suggestions which led us to refine many aspects of the work and improve the overall quality of the study.\\\vspace{-2mm}

\textbf{Software:} ArielRad \citep[v2.4.16,][]{mugnai_ar}, TauREx3 \citep{al-refaie_taurex3}, TauREx3 ACE plug-in \citep{al-refaie_taurex3_chem,venot_chem}, Astropy \citep{astropy}, h5py \citep{hdf5_collette}, Matplotlib \citep{Hunter_matplotlib}, Multinest \citep{Feroz_multinest,buchner_multinest}, Pandas \citep{mckinney_pandas,reback_pandas}, Numpy \citep{oliphant_numpy}, SciPy \citep{scipy}, corner \citep{corner}.\\ \vspace{-2mm}

\textbf{Linelists:} H$_2$O \citep{polyansky_h2o}, CH$_4$ \citep{exomol_ch4}, CO \citep{li_co_2015}, CO$_2$ \citep{rothman_hitremp_2010}, CIA \citep{abel_h2-h2, fletcher_h2-h2, abel_h2-he}.  \\

\clearpage

\begin{longtable*}{lllllllll}
\caption{Currently-known exoplanets which are considered here to be potential targets for Ariel. The list will continue to evolve as surveys discover more planets.}\\ \hline
Planet Name & Tier & Method & Planet Name & Tier & Method & Planet Name & Tier & Method\\ \hline
\endfirsthead

\multicolumn{9}{c}%
        {{\bfseries Table \thetable\ Continued from previous page}} \\
        \hline
        
Planet Name & Tier & Method & Planet Name & Tier & Method & Planet Name & Tier & Method  \\ \hline
 \endhead

 \hline
 \endfoot
55 Cnc e & 2 & Transit	&	HAT-P-30 b & 3 & Eclipse	&	HATS-29 b & 1 & Eclipse	\\
AU Mic b & 3 & Transit	&	HAT-P-31 b & 1 & Eclipse	&	HATS-3 b & 2 & Eclipse	\\
AU Mic c & 2 & Transit	&	HAT-P-32 b & 3 & Eclipse	&	HATS-30 b & 2 & Eclipse	\\
CoRoT-1 b & 1 & Eclipse	&	HAT-P-33 b & 3 & Eclipse	&	HATS-31 b & 2 & Eclipse	\\
CoRoT-11 b & 1 & Eclipse	&	HAT-P-34 b & 2 & Eclipse	&	HATS-33 b & 2 & Eclipse	\\
CoRoT-2 b & 3 & Eclipse	&	HAT-P-35 b & 2 & Eclipse	&	HATS-34 b & 1 & Eclipse	\\
DS Tuc A b & 1 & Eclipse	&	HAT-P-36 b & 2 & Eclipse	&	HATS-35 b & 2 & Eclipse	\\
EPIC 211945201 b & 1 & Transit	&	HAT-P-37 b & 1 & Eclipse	&	HATS-37 A b & 2 & Transit	\\
\end{longtable*}
 
\null\clearpage

\begin{longtable*}{llllllllllll}
\caption{TESS planet candidates which are considered here to be potential targets for Ariel. The list will continue to evolve as TESS discovers more potential planets.}\\ \hline
TOI & Tier & Method & TOI & Tier & Method & TOI & Tier & Method & TOI & Tier & Method\\ \hline
\endfirsthead

\multicolumn{12}{c}%
        {{\bfseries Table \thetable\ Continued from previous page}} \\
        \hline
        
TOI & Tier & Method & TOI & Tier & Method & TOI & Tier & Method  & TOI & Tier & Method\\ \hline
 \endhead
  \hline
 \endfoot
 
139.01 & 1 & Transit	&	375.01 & 1 & Eclipse	&	603.01 & 2 & Transit	&	778.01 & 3 & Eclipse	\\
145.01 & 1 & Transit	&	376.01 & 1 & Transit	&	610.01 & 1 & Transit	&	783.01 & 1 & Transit	\\
148.01 & 1 & Transit	&	383.01 & 1 & Transit	&	611.01 & 1 & Transit	&	784.01 & 1 & Transit	\\
153.01 & 1 & Transit	&	392.01 & 3 & Eclipse	&	614.01 & 1 & Eclipse	&	790.01 & 1 & Transit	\\
155.01 & 1 & Transit	&	393.01 & 1 & Transit	&	615.01 & 3 & Eclipse	&	791.01 & 1 & Transit	\\
159.01 & 3 & Eclipse	&	399.01 & 1 & Transit	&	618.01 & 1 & Transit	&	802.01 & 1 & Transit	\\
166.01 & 2 & Transit	&	406.01 & 1 & Transit	&	620.01 & 1 & Transit	&	811.01 & 1 & Transit	\\
170.01 & 1 & Eclipse	&	412.01 & 3 & Eclipse	&	621.01 & 3 & Eclipse	&	812.01 & 1 & Transit	\\
\end{longtable*}

\bibliography{main}

\end{document}